\documentclass[11pt]{article}

\usepackage{amsmath, amssymb}
\usepackage{microtype} 
\usepackage {multirow} 
\usepackage{booktabs} 
\usepackage{array}
\usepackage{authblk} 
\usepackage{graphicx}
\usepackage{cite}
\usepackage{color}
\usepackage{geometry} 
\usepackage[labelfont=bf,labelsep=period]{caption}
\usepackage{xspace} 
\usepackage{hyperref}

\makeatletter
\renewcommand{\@biblabel}[1]{\quad#1.}
\makeatother

\date{}

\newcommand{\psd}{$\mathrm{PSD95}^{\mathrm{pdz3}}$}

\newcommand{\es}[2]{\begin{equation}\label{#1}\begin{split}#2\end{split}\end{equation}}

\newcommand{\abs}[1]{\lvert #1 \rvert}

\newcommand{\hhblits}{{\sffamily HHblits}\xspace}
\newcommand{\pfam}{{\sffamily Pfam}\xspace}
\newcommand{\arxiv}{{\sffamily arXiv}\xspace}

\DeclareMathOperator{\MI}{MI}
\DeclareMathOperator*{\argmax}{arg\,max}
\DeclareMathOperator*{\Tr}{Tr}
\DeclareMathOperator{\iqr}{IQR}

\begin{document}

\title{Protein sectors: statistical coupling analysis versus conservation}

\author[1,2]{Tiberiu Te\c sileanu\footnote{e-mail: ttesileanu@gmail.com}}
\author[1,3]{Lucy J.\ Colwell}
\author[1,4]{Stanislas Leibler}
\affil[1]{The Simons Center for Systems Biology and The School of Natural Sciences, Institute for Advanced Study, Einstein Drive, Princeton, NJ 08540, USA}
\affil[2]{Initiative for the Theoretical Sciences, CUNY Graduate Center, 365 Fifth Avenue, New York, NY 10016, USA}
\affil[3]{Department of Chemistry, University of Cambridge, Lensfield Road, Cambridge CB3 0WA, UK}
\affil[4]{Center for Studies in Physics and Biology and Laboratory of Living Matter, The Rockefeller University, 1230 York Avenue, New York, NY 10065, USA}

\maketitle

\begin{abstract}
Statistical coupling analysis (SCA) is a method for analyzing multiple sequence alignments that was used to identify groups of coevolving residues termed ``sectors''. The method applies spectral analysis to a matrix obtained by combining correlation information with sequence conservation. It has been asserted that the protein sectors identified by SCA are functionally significant, with different sectors controlling different biochemical properties of the protein. Here we reconsider the available experimental data and note that it involves almost exclusively proteins with a single sector. We show that in this case sequence conservation is the dominating factor in SCA, and can alone be used to make statistically equivalent functional predictions.
Therefore, we suggest shifting the experimental focus to proteins for which SCA identifies several sectors. Correlations in protein alignments, which have been shown to be informative in a number of independent studies, would then be less dominated by sequence conservation.
\end{abstract}

\newpage
\tableofcontents
\newpage

%

\section{Introduction}

A fundamental question in biology is the relation between the amino acid sequence of a protein and its function and three-dimensional structure. Given the rapid growth in the sequence data available from many organisms, it has become possible to use statistical sequence analysis to approach this question. Based on sequence similarity, protein sequences can be grouped into families thought to share common ancestry; the proteins in such a family typically perform related functions and fold into similar structures~\cite{Do2008, Notredame2002}. It has been shown in many studies that a statistical analysis of a multiple sequence alignment (MSA) corresponding to a given protein family can be used to find amino acids that control different aspects of a protein's function or structure.

A basic statistical quantity that can be calculated for a multiple sequence alignment is the distribution of amino acids at each site. In particular, the level of sequence conservation at each site is of biological relevance, since it is expected that conservation is low in the absence of selective pressures. For this reason, conservation has long been used to predict which parts of a protein are most likely to be functionally significant~\cite{Schneider1986, Zvelebil1987, Hollstein1991, Cargill1999, Ng2003}.

More recently, the availability of large sets of protein sequences has made it possible to also estimate higher-order statistics, such as the correlations between the amino acids found at each pair of sequence positions. In a number of examples, these statistics have been shown to contain information about the structure and function of proteins~\cite{Russ2005, Socolich2005, Marks, Morcos2011, Weigt2009}. One way in which pairwise correlations might arise is for a deleterious mutation at a given position to be compensated by a mutation at a different position. This can yield a scenario in which the two individual mutations are relatively rare, but the combination of both is common in natural proteins.

Statistical coupling analysis (SCA) was introduced by Lockless and Ranganathan in 1999 as a way to infer energetic interactions within a protein from a statistical analysis of a multiple sequence alignment~\cite{Lockless1999}. The authors compared the statistics of an alignment of PDZ domain sequences to measurements of the binding affinity between a particular member of the alignment (\psd) and its cognate ligand. The statistical analysis assumed that the frequencies of mutations obey a Boltzmann distribution as a function of binding free energy, allowing estimation of the binding affinity by $\Delta G_i \sim \log f_i$, where $f_i$ is the frequency of an amino acid type at a given site in the alignment. By conditioning on amino acid type at a second site, they calculated the amount by which the effect of a mutation at one site changed depending on the amino acid present at the second site: $\Delta \Delta G_{i|j} = \Delta G_{i|j} - \Delta G_i \equiv \Delta \Delta G^{\text{stat}}$. This gave an estimate for the effective coupling between the sites.

The assumptions behind the original formulation of SCA are likely to be violated, since the selective pressures acting on a protein are more complex than simply maximizing binding to a ligand. Despite this, the method seemed to be effective. In the original paper~\cite{Lockless1999}, mutant cycle analysis was used to measure $\Delta \Delta G^{\text{binding}}$, the amount by which the effect of a given mutation on ligand binding affinity of \psd~changes when the mutation occurs on a background containing a second mutation. This can be written as $\Delta\Delta G^{\text{binding}} = \Delta G_{i|j} - \Delta G_i$, where now $\Delta G$ represents a change in the physical free energy as opposed to a statistical construct. The quantity $\Delta \Delta G^{\text{binding}}$ was observed to be well correlated with the statistically-calculated $\Delta \Delta G^{\text{stat}}$. The set of residues identified by SCA to be coupled with a particular site known to be important for binding specificity of the PDZ domain was found to physically connect distal functional sites of the protein~\cite{Lockless1999}. This led to the suggestion that these residues may mediate an allosteric response. Experimental evidence later showed that indeed some of the residues identified by SCA participate in allostery~\cite{Fuentes2004, Peterson2004, Suel2003, Hatley2003}. Moreover, in a different study, a large fraction of the artificial WW domains built by conserving the pattern of statistical couplings calculated by SCA were observed to be functional, while sequences built to conserve single-site statistics alone were not~\cite{Russ2005, Socolich2005}.

Motivated by these observations, Halabi et al.\ reformulated SCA in purely statistical terms, avoiding the assumptions related to energetics~\cite{Halabi2009}. The reformulation amounted to a particular way of combining correlations with conservation. The basic idea was to multiply each element of the covariance matrix $C_{ij}$ by a product $\phi_i \phi_j$, yielding the ``SCA matrix'' $\tilde C_{ij} = \phi_i \phi_j C_{ij}$. The ``positional weights'' $\phi_i$ were a function of the frequency $f_i$ of the most prevalent amino acid at each position, and were roughly given by $\phi_i \sim \log [f_i / (1 - f_i)]$. This particular form was chosen to reproduce the results from the original formulation of SCA~\cite{Halabi2009, Ranganathan2011url}. In subsequent work regarding SCA, several variations on this basic idea were used; all of these yield similar though not identical results and are described more precisely in sections~\ref{sec:methods_sca} and~\ref{sec:supp_sca}.

Running either the original or the reformulated analysis on several examples~\cite{Lockless1999, Suel2003, Hatley2003, Russ2005, Socolich2005, Halabi2009, Smock2010}, it was noticed that the resulting SCA matrix had an approximate block structure. In analogy to previous work in finance, Halabi et al.\ analyzed this structure by looking at the top eigenvectors of the SCA matrix~\cite{Halabi2009, Bouchaud2009}. The corresponding groups of residues were called ``protein sectors'' because similar clusters observed in the correlations of stock prices were found to correspond to financial sectors. Experiments found that mutating residues in distinct sectors specifically affected different phenotypes of the protein~\cite{Halabi2009}, leading to the suggestion that each SCA sector might comprise a group of amino acids that control a particular phenotype.

\begin{table*}
	\caption{\textbf{The different meanings that can be associated with protein sectors}}
	\centering
	\begin{tabular}{p{0.15\hsize}p{0.34\hsize}p{0.40\hsize}}
			\hline
			Interpretation of sector & Signature & Possible ways of quantitative exploration\\
			\hline
			statistical & clusters of correlated mutations in MSA & statistical analysis of MSA\\
			evolutionary & maintains identity under evolutionary dynamics & artificial evolution experiments\\
			structural & distinct physical properties compared to surroundings & NMR, room-temperature X-ray crystallography, MD simulations\\
			functional & altering sector positions changes functional properties & mutagenesis studies\\
			\hline
	\end{tabular}
	\label{tab:secdiffmeanings}
\end{table*}

It is important to note that there are several subtly different meanings that have been attributed to protein sectors (see Table~\ref{tab:secdiffmeanings}). The description outlined above defines protein sectors as the results of a statistical analysis of a multiple sequence alignment. This definition depends on the statistical method employed; it would, for example, depend on the choice of positional weights in the case of SCA, or on the precise thresholds and methods used for clustering. To distinguish this from other meanings, we will call these \emph{statistical sectors} (or \emph{SCA sectors} when the statistical method is SCA).

The sectors identified by SCA have also been given an \emph{evolutionary} interpretation~\cite{Halabi2009, Smock2010, Rivoire2012}, based on the fact that they are defined as groups of residues whose mutations are correlated in an alignment, and the sequences in the alignment are likely to be evolutionarily related. However, this argument is insufficient to prove the evolutionary nature of the statistical sectors, given that their precise composition is dependent on the statistical method employed~\cite{Colwell2014}. Thus it is difficult to assess to what extent the sector's composition is actually related to the evolutionary process itself, as opposed to the choice of the statistical method. Strikingly, Halabi et al.\ showed that for an alignment of serine proteases, one of the sectors can be used to distinguish between vertebrates and invertebrates, suggesting that indeed an evolutionary interpretation may be appropriate~\cite{Halabi2009}. However, before concluding that in general SCA sectors have an evolutionary interpretation, it would be important to extend these studies to different alignments. An alternative, more direct, approach would be to perform artificial evolution experiments to check whether the SCA sectors maintain their integrity under strong selection, or whether new sectors can be created in this way. In addition, such experiments would provide data on the evolutionary dynamics of proteins, and thus help to define more precisely the notion of evolutionary sectors.

Another surprising property of the groups of residues identified by SCA is that they usually form contiguous structures in the folded protein, although they are not contiguous in sequence~\cite{Halabi2009, Smock2010, Reynolds2011, McLaughlinJr2012, Rivoire2012}. This suggests the notion of \emph{structural sectors}, groups of residues having different physical properties compared to their surroundings. An experimental test for such inhomogeneities inside proteins could employ NMR spectroscopy to follow the dynamics of specific atoms while the protein is undergoing conformational change~\cite{Kern2003, Fuentes2004}. In addition, analyzing room-temperature X-ray diffraction data could shed light on residues with coupled mobility or increased fluctuations in an ensemble of structures~\cite{Matoba2003, Fraser2009} (Doeke Hekstra, personal communication). Alternatively, this kind of experiments could be done \emph{in silico} using for example molecular dynamics simulations to identify correlated motions in the protein~\cite{Dhulesia2008} (Olivier Rivoire, personal communication).

Finally, as mentioned above, a number of mutational studies have suggested yet another interpretation of the sectors identified by SCA as \emph{functional sectors}, groups of amino acids that cooperate to control certain phenotypic traits of a protein, such as binding affinity~\cite{Lockless1999, Halabi2009, Smock2010, McLaughlinJr2012}, denaturation temperature~\cite{Halabi2009}, or allosteric behavior~\cite{Suel2003, Hatley2003, Fuentes2004, Peterson2004, Reynolds2011}. It is this aspect of the sectors that has been most emphasized in the literature.

In the language we just introduced, we can say that there is some data suggesting that SCA can identify groups of residues that act as evolutionary, structural, and functional sectors in a protein. It is important to note that these aspects can exist independently of one another. As an example, the existence of a physical inhomogeneity overlapping the statistical sector positions would support the idea that SCA can identify structural sectors, but would provide no guarantee that these also have an associated phenotype. For this reason, independent experimental verification is needed to support each of these claims.

We focus here on the experimental evidence supporting the hypothesis that SCA sectors act as functional sectors of proteins~\cite{Russ2005, Socolich2005, Halabi2009, Smock2010, Reynolds2011, McLaughlinJr2012}. We note that with the exception of Halabi et al.~\cite{Halabi2009}, this data refers to proteins in which a single SCA sector was identified, and we show that in this case, within statistical uncertainties, a method based on sequence conservation can identify functional residues as well as SCA. We also give a simple mathematical argument describing why this might happen.

Given that conservation information is explicitly used in calculating the SCA matrix, it is not surprising that SCA sectors are related to conservation. However, what we show here is that conservation dominates the SCA calculations in the single-sector case; thus, in order to establish whether the functional significance of SCA sectors is more than what is expected from single-site statistics alone, experiments need to focus on the examples where SCA identifies several sectors. The analysis of serine proteases described above provides such a study~\cite{Halabi2009}, but it is essential to have more data for different protein families to assess the robustness and generality of these observations.

\section{Results}
\begin{figure}[!t]
	\centering
	\includegraphics[width=\textwidth]{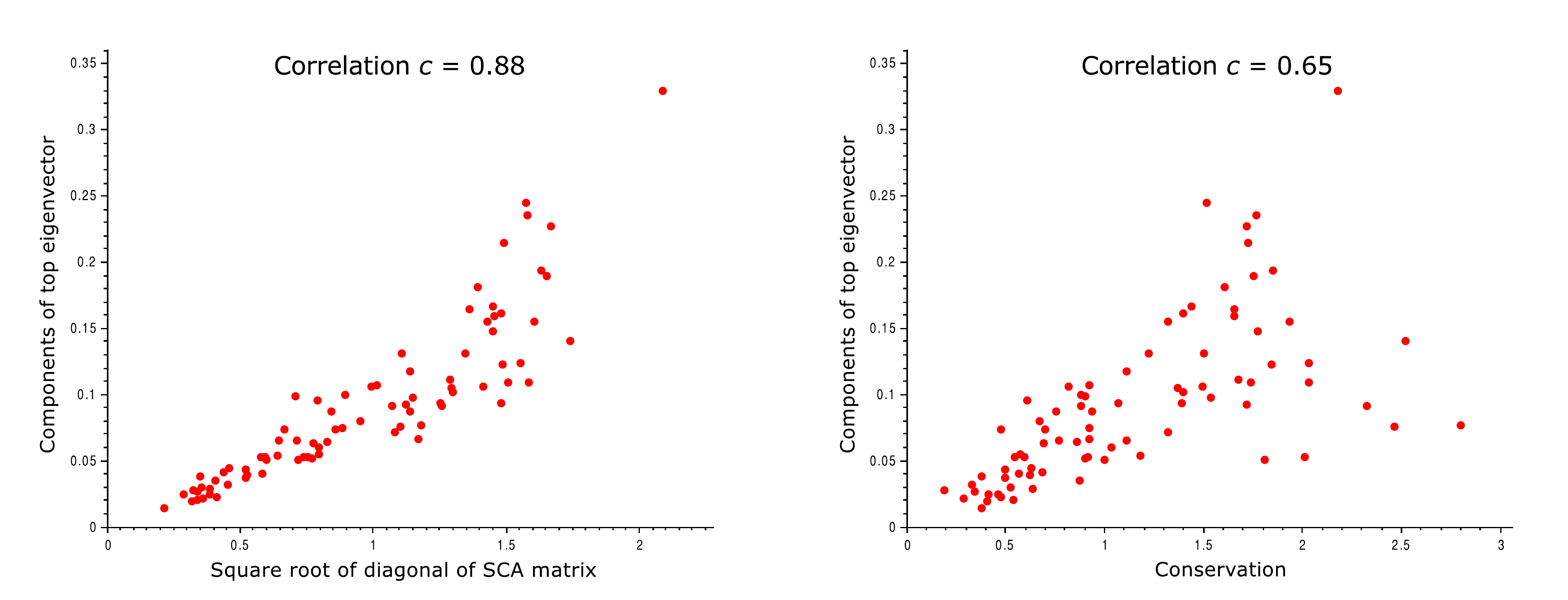}
	\caption{\textbf{Relation between the components of the top eigenvector of the SCA matrix and the square root of its diagonal elements \emph{(left)} or conservation \emph{(right)}.} This was obtained for the PDZ alignment, but the results are similar for other alignments.}
	\label{fig:pdz_topev_vs_diag_and_cons}
\end{figure}

We analyze existing experimental datasets to compare the functional significance of SCA residues to that of conserved residues. Some of these datasets (PDZ, DHFR, and the voltage-sensing domains of potassium channels) have already been analyzed using SCA; one of them (\textit{lacI}) has not. We show that in all these cases, conservation identifies functional positions just as effectively as SCA. This holds true for a wide range of choices of thresholds used to define conserved and functional residues, respectively.

There have been several versions of the SCA approach that have been used in the literature. Indeed, each of the three datasets mentioned above for which SCA has been applied was analyzed using a different variation of the method. To avoid ambiguities, here we use a uniform method for all the alignments (see Methods for details). While Halabi et al.\ explicitly ignored the top eigenvector based on an analogy to finance~\cite{Halabi2009}, here we focus only on the top eigenvector. The reason for this is that, besides Halabi et al., all other published studies related to SCA have included this mode in their analysis~\cite{Smock2010, Reynolds2011, McLaughlinJr2012}.\footnote{In the case of DHFR, Reynolds et al.\ define the single sector using not only the top eigenvector, but the top five~\cite{Reynolds2011}. As we discuss below, the results from that paper are, however, not significantly changed if the sector is defined based only on the top eigenvector.} The alignments used were generated using the \hhblits software~\cite{Remmert2012} with a consistent set of options (see Methods for details).

The definition of conservation we use is based on the relative entropy (Kullback-Leibler divergence)~\cite{Halabi2009},
\begin{equation}
	\label{eq:defConservation0}
	D_i = \sum_a f_i(a) \log \frac {f_i(a)} {q(a)} \,,
\end{equation}
where $f_i(a)$ is the frequency at which amino acid $a$ occurs in column $i$ of the multiple sequence alignment and $q(a)$ is the background frequency for amino acid $a$. We use the same background frequencies as employed in SCA, which were calculated by Lockless and Ranganathan by averaging over a large protein database~\cite{Lockless1999}. Other common definitions for conservation, such as the frequency of the most prevalent amino acid at a given position, tend to be highly correlated with $D_i$ described above.

It is important to note that the qualitative results are unchanged regardless which version of SCA, which definition for conservation, or which alignments are used (see appendices for details). This work was motivated by the empirical observation that for many alignments the components of the top eigenvector correlate strongly with the diagonal elements of the SCA matrix (see Figure~\ref{fig:pdz_topev_vs_diag_and_cons}, left). The values of the diagonal elements can be calculated in terms of single-site statistics, raising the question whether correlations are needed to find the positions comprising the top sector. In fact, the components of the top eigenvector are also well-correlated with the conservation $D_i$ defined above (see Figure~\ref{fig:pdz_topev_vs_diag_and_cons}, right). Note that these observations are not particularly surprising, given that the SCA matrix is weighted by quantities related to conservation. However, they also raise the question whether the observed functional significance of SCA sectors~\cite{Lee2009, Reynolds2011, McLaughlinJr2012} could be due to conservation instead of correlations.

\subsection{PDZ domains}
The ability of SCA to identify residues that are important for protein function was recently tested in a high-throughput experiment involving a PDZ domain~\cite{McLaughlinJr2012}. Each amino acid of the \psd~domain was mutated to all 19 alternatives and the binding affinity of the resulting mutants to the \psd\ cognate ligand was measured. The measurement involved a bacterial two-hybrid system in which the PDZ domain was fused to the DNA-binding domain of the $\lambda$-cI repressor, while the ligand was fused to the $\alpha$ subunit of the \textit{E.~coli} RNA polymerase. This was used to control expression of GFP, which allowed the binding affinity between \psd~and its ligand to be estimated using fluorescence-activated cell sorting (FACS). In order to quantify the sensitivity to mutations at a given site, the mutational effects on binding affinity were averaged over all 20 possible amino acids at that site. While mutations at most sites were found to have a negligible effect on ligand binding, 20 sites were identified where mutations had a significant deleterious effect~\cite{McLaughlinJr2012}.

\begin{figure}
	\centering
	\includegraphics[width=0.6\textwidth]{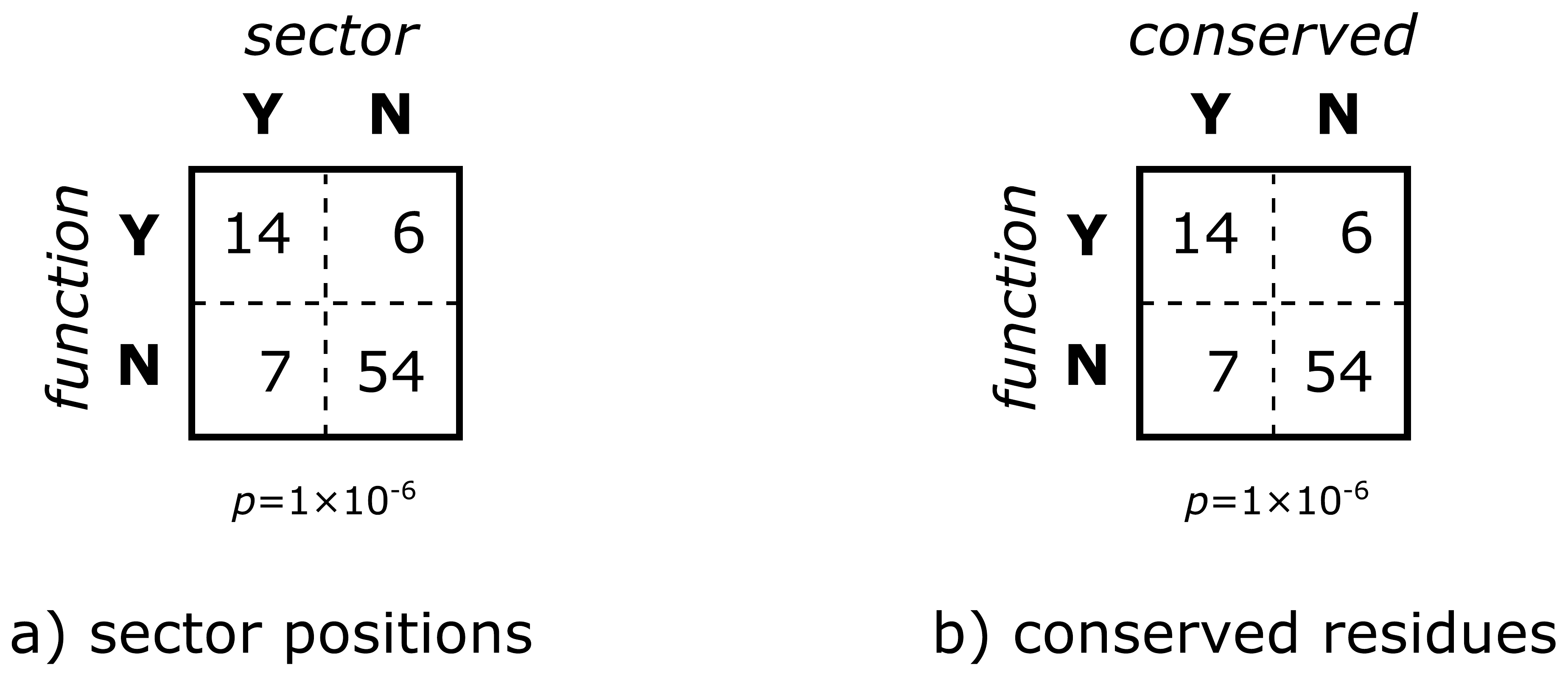}
	\caption{\textbf{Contingency tables testing whether a \psd\ residue belonging to a sector or being highly-conserved is associated with significant functional effect upon mutagenesis.} The tables are identical although only 57\% of the residues are shared between the sector and the conserved positions.}
	\label{fig:pdz_fisher}
\end{figure}

The sector identified by SCA according to the methodology outlined above was found to indeed contain residues that are more likely to have functional significance than randomly chosen positions in the protein: 14 of the 21 sector residues are functionally significant, or 67\%, compared to 25\% for the entire protein (see Figure~\ref{fig:pdz_fisher}a). This is statistically-significant according to a Fisher exact test (one-tailed $p = 1\times 10^{-6}$), and this result is robust to changing the threshold used to define the sector. This mirrors the results from McLaughlin Jr.\ et al.~\cite{McLaughlinJr2012}, obtained there with a different alignment constructed using a structural alignment algorithm~\cite{Lockless1999}.

\begin{figure}
	\centering
	\includegraphics[width=0.6\textwidth]{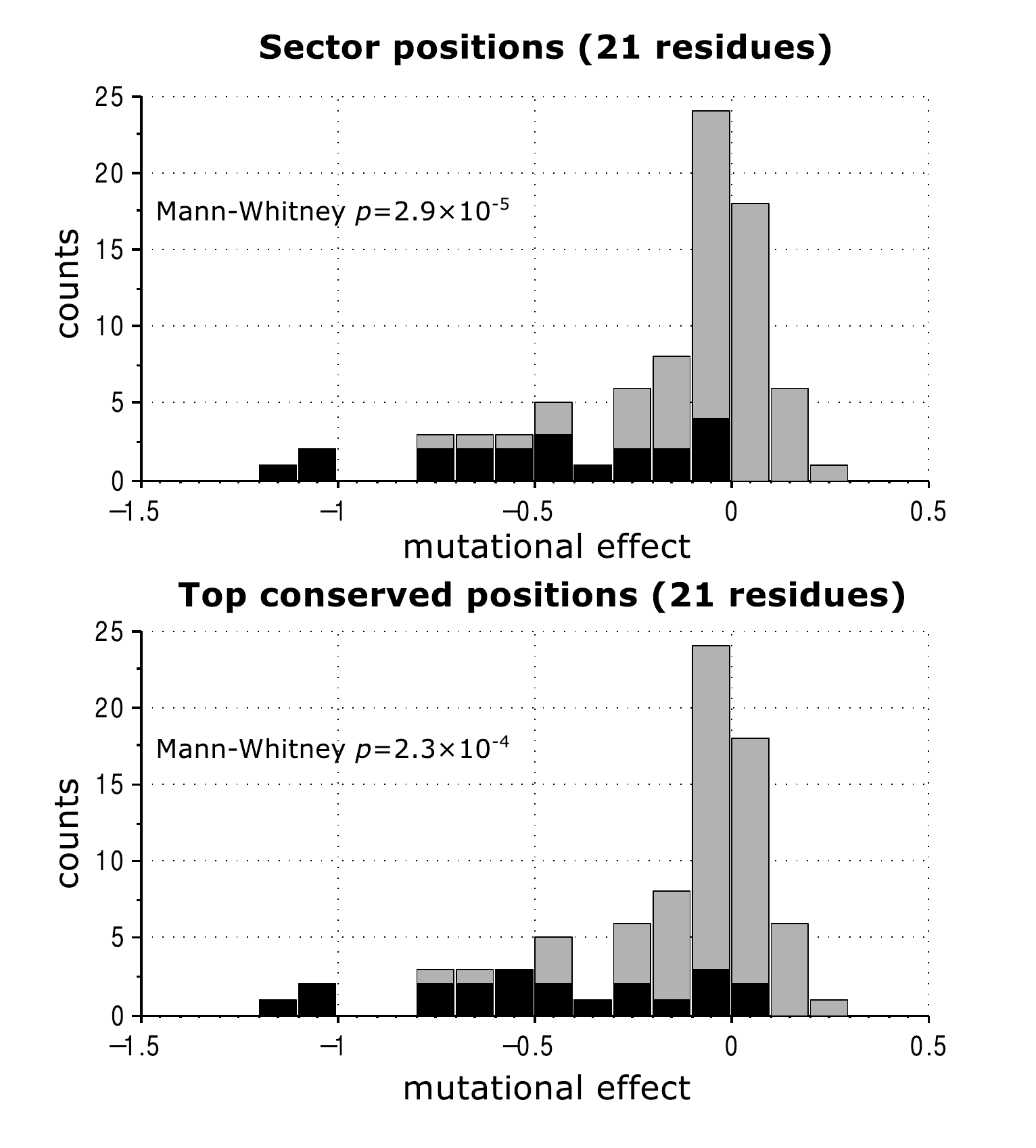}
	\caption{\textbf{Histograms showing the effect of mutations on binding affinity of \psd~with cognate ligand, for all mutations (gray), and for mutations to selected positions (black).} Each of the histograms in black contains 21 positions, with the largest SCA scores (top), or the largest conservation levels (bottom). A Mann-Whitney $U$ test cannot find a statistically-significant difference between the distribution of mutational effects for sector positions and the one for conserved positions ($p = 0.9$). The mutational effect $\langle \Delta E_i^x\rangle_x$ is a dimensionless quantity calculated as in McLaughlin Jr. et al.~\cite{McLaughlinJr2012}.}
	\label{fig:pdz_hist_sca_cons}
\end{figure}

There is another way of assessing the functional relevance of the sector positions that avoids making a sharp distinction between functional and non-functional residues~\cite{McLaughlinJr2012}. The functional effects of mutations at all the positions in the domain were used to define a background distribution showing how likely an effect of a given magnitude was. If the sector is able to identify functionally-relevant positions, then the distribution of functional effects restricted to the sector positions should differ from this background distribution. Figure~\ref{fig:pdz_hist_sca_cons} (top) shows the comparison for the PDZ experiment. A two-sample Mann-Whitney $U$ test~\cite{Mann1947} finds that indeed sector positions have a statistically-significant distribution of functional effects compared to all residues.\smallskip

We now test whether we could have obtained similar results by considering only sequence conservation. Indeed, although the 21 most conserved residues are different from the 21 residues identified by SCA (only about 60\% are shared), the fraction of these residues that is functionally significant is the same (see Figure~\ref{fig:pdz_fisher}b). The histogram of functional effects is also essentially the same between SCA sector residues and conserved residues (see Figure~\ref{fig:pdz_hist_sca_cons}, bottom), and in fact a Mann-Whitney $U$ test confirms that the difference is not statistically significant.

McLaughlin Jr.\ et al.\ performed a similar analysis and obtained similar histograms as our Figure~\ref{fig:pdz_hist_sca_cons} (see Figure 3a in their paper~\cite{McLaughlinJr2012}). The reason why our results seem so different is that, due to an error, the top histogram in Figure 3a in McLaughlin Jr.\ et al. is missing the data for the five sector residues that do not have a significant mutational effect. These five sector residues are mentioned and taken into account in other parts of the paper by McLaughlin Jr.\ et al~\cite{McLaughlinJr2012}, for example in Table S6a in the supplementary information, but they do not appear in the histogram. Had they been included, the histograms for conserved residues and that for SCA sector residues would look almost identical, in agreement with our results.

We stress again that these results do not imply that correlations in protein alignments are not informative. Indeed, as mentioned in the introduction, experimental data on the creation of artificial WW domains showed that ignoring correlations leads to non-functional proteins, while proteins designed based on conservation-weighted correlations can often be functional~\cite{Russ2005}. Moreover, correlation information was used to provide quite accurate predictions for contact maps and three-dimensional structures of a variety of proteins~\cite{Weigt2009, Morcos2011, Marks}. This is not possible using single-site statistics alone. The question we are asking, however, is whether the particular way in which alignment correlations are used in SCA is more useful for predicting functional information than conservation. The answer seems to be negative for the case of PDZ.

\begin{figure}
	\centering
	\includegraphics[width=0.7\textwidth]{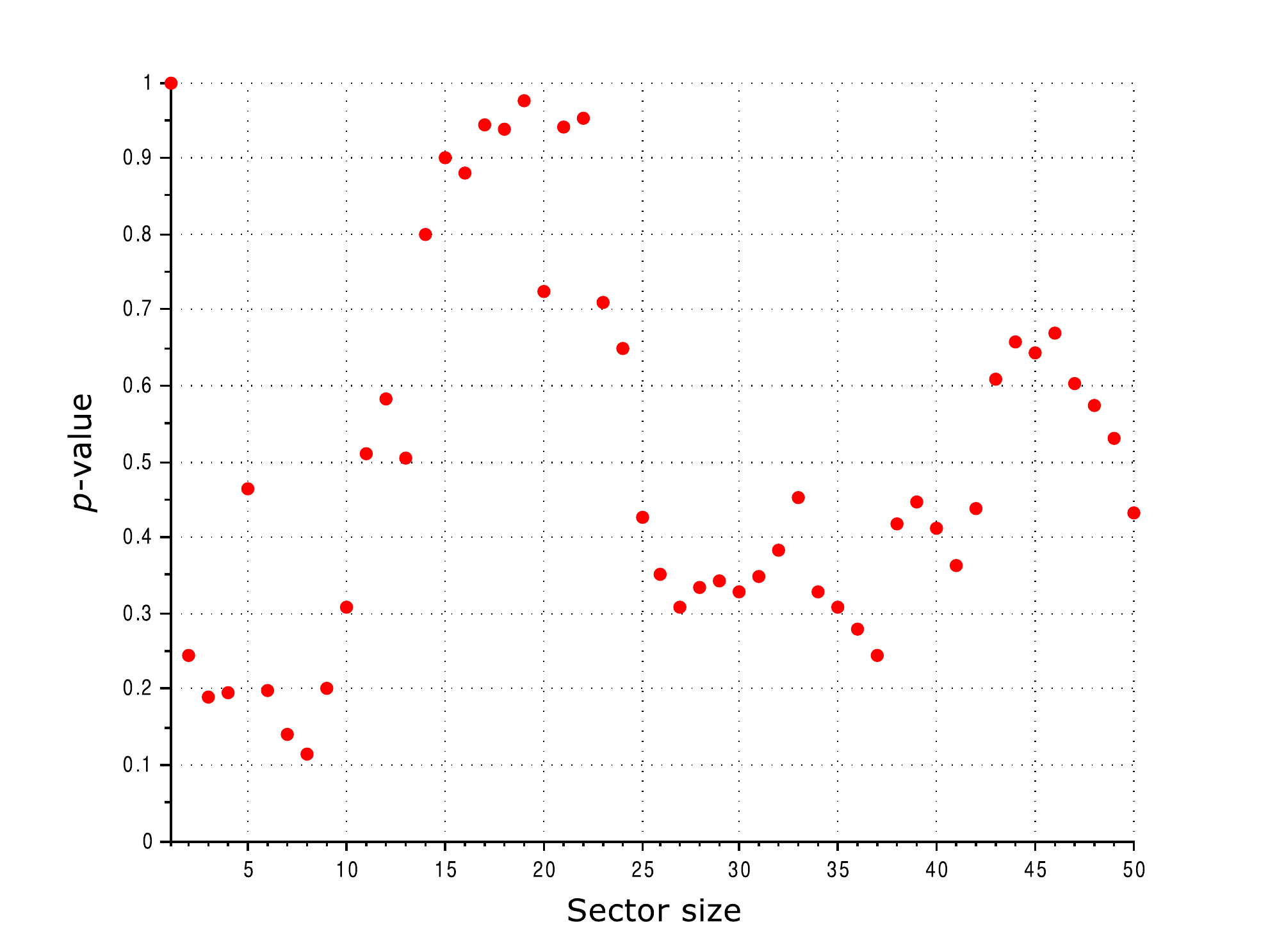}
	\caption{\textbf{Comparison of the ability of the SCA sector and conservation to predict the functional effect of mutation of \psd~residues for various sector sizes.} The vertical axis shows the $p$-value for a two-sample, two-tailed Mann-Whitney $U$ test comparing the distribution of mutational effects for sector residues \textit{vs.}~conserved residues.}
	\label{fig:pdz_secs_vs_cons_sweep}
\end{figure}

All the observations reported above are qualitatively the same when using different alignments, including the alignment employed by McLaughlin Jr.\ et al.~\cite{McLaughlinJr2012} and a \pfam alignment. The observations are also robust to varying the threshold used for defining the sector: in Figure~\ref{fig:pdz_secs_vs_cons_sweep} we show a statistical comparison between the SCA sector and conserved residues calculated for various sizes of the sector.

Note that there are some potential caveats for the statistical tests we used. One assumption of both the Mann-Whitney $U$ test and the $\chi^2$ test employed above is that the samples analyzed are independent. In our case, the samples are the mutational effects at different residues in a protein domain, which are unlikely to be independent. Designing a statistical test that overcomes this difficulty would require a detailed model of evolutionary dynamics that accurately describes the relation between the binding affinity of \psd~to its cognate ligand, and the evolutionary information contained in a multiple sequence alignment. To our knowledge, there is unfortunately no unambiguous way of constructing such a model. Despite these issues, the analysis presented here suggests that, for the top sector, SCA is not significantly better than conservation at predicting functionally-important sites.

\subsection{Dihydrofolate reductase}
\begin{figure}
	\centering
	\includegraphics[width=0.6\textwidth]{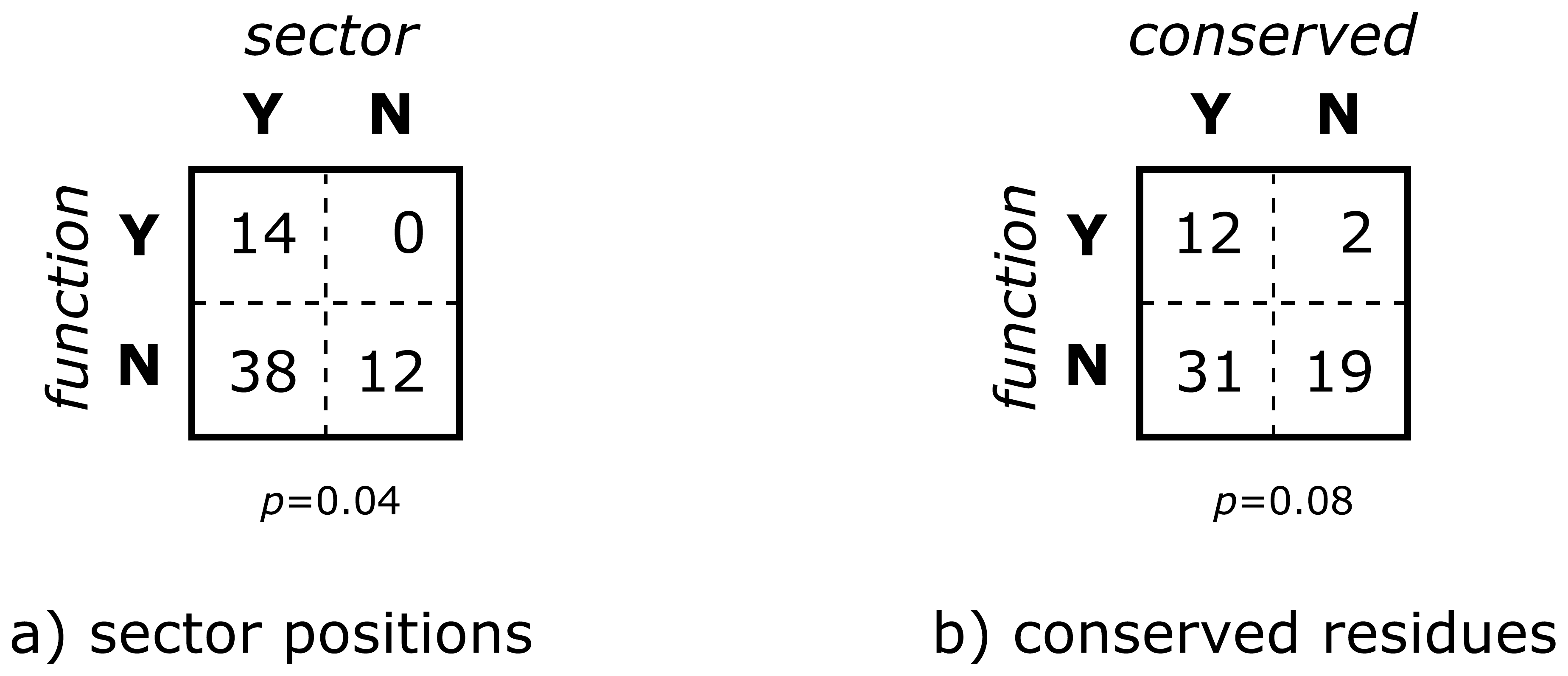}
	\caption{\textbf{Contingency tables testing whether sector residues or conserved residues are more likely to ``touch'' functionally-significant LOV2 insertion points for the DHFR protein analyzed in Reynolds et al.~\cite{Reynolds2011}.} A $\chi^2$ test cannot reject the hypothesis that the two contingency tables are drawn from the same distribution ($p = 0.2$).}
	\label{fig:dhfr_fisher}
\end{figure}

The case of dihydrofolate reductase (DHFR)~\cite{Reynolds2011} exhibits some interesting differences from PDZ. The experimental assay in this case involved perturbing the DHFR protein by attaching a light-sensitive domain (LOV2) between the atoms of the peptide bond immediately preceding each surface residue. The experiment used a folate auxotroph mutant of \textit{E.~coli} whose growth was rescued by a plasmid containing DHFR and thymidylate synthetase genes. The growth rate of the bacteria, which was measured with a high-throughput sequencing method, was shown to be approximately proportional to the catalytic efficiency of DHFR. The functional effect of each insertion of the LOV2 domain was measured by the difference in growth rates between lit and dark conditions. Out of the 61 measured surface sites, 14 were found to have a significant functional effect~\cite{Reynolds2011}.

The effects of the insertion of the LOV2 domain are not localized on a single residue of the protein, which makes the analysis of the functional significance of the SCA sector positions more complicated in the case of DHFR. We follow here the method employed in the original study by Reynolds et al., which is to define a range around the insertion point within which a residue could conceivably feel the influence of the inserted domain~\cite{Reynolds2011}. More specifically, $4$ \AA\ spheres were centered on each of the four atoms forming the peptide bond broken by the insertion of LOV2, and any residues having at least one atom centered within any of these spheres was counted as ``touching'' the light-sensitive residue. The exact size of the cutoff is not important: we repeated the analysis with the cutoff set to $3$ \AA\ and $5$ \AA\ and obtained the same qualitative results.

Using the methodology described above, the SCA sector identified from the top eigenvector of the SCA matrix is found to ``touch'' all 14 of the functionally-significant LOV2 insertion sites. A set of conserved residues of the same size as the SCA sector ``touches'' 12 of the functionally-significant sites, and the difference is not statistically significant (see Figure~\ref{fig:dhfr_fisher}).

\begin{figure}
	\centering
	\includegraphics[width=0.7\textwidth]{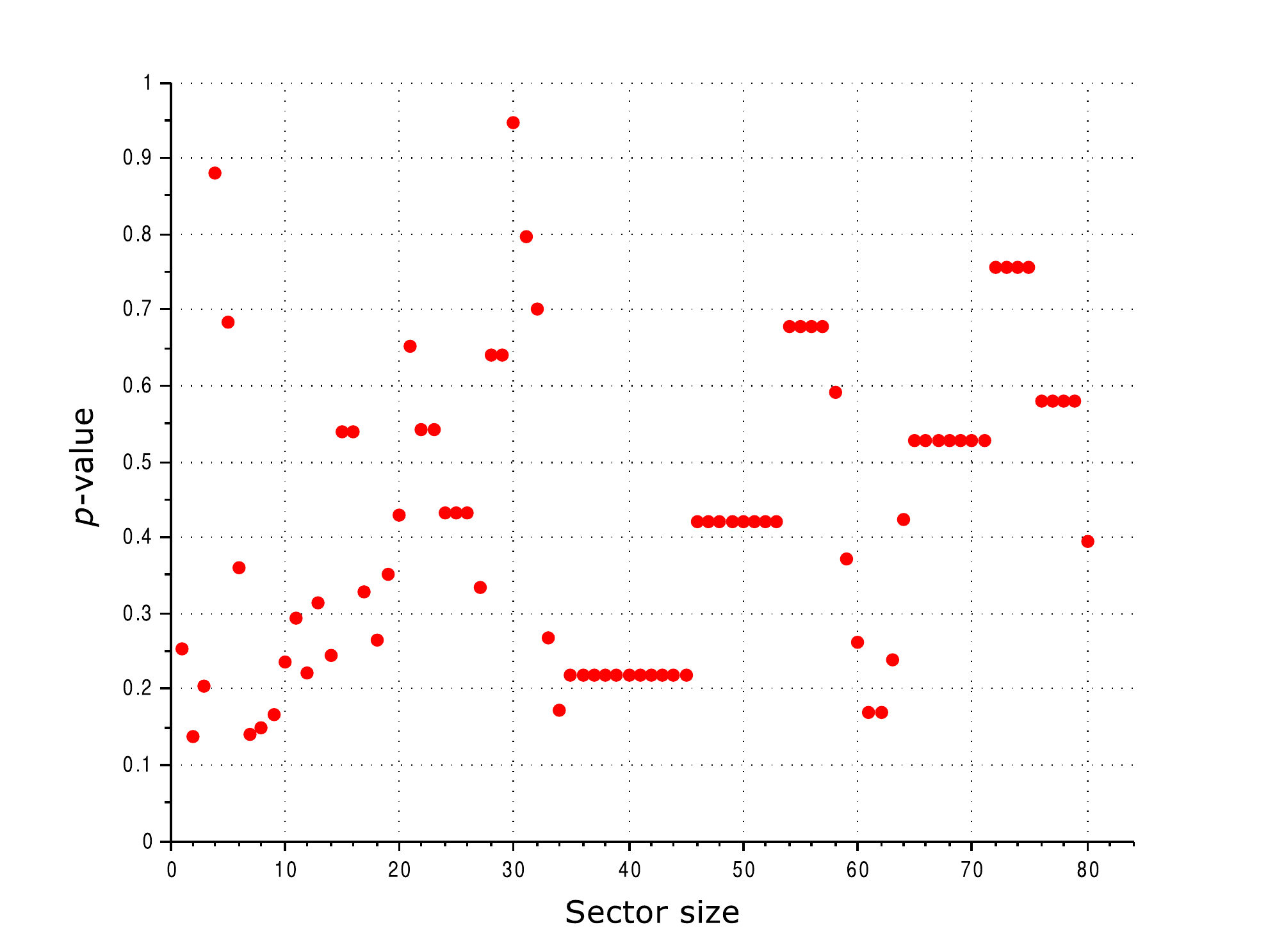}
	\caption{\textbf{Comparison of the ability of the SCA sector and conserved residues to ``touch'' the functionally-significant sites of DHFR identified by Reynolds et al.~\cite{Reynolds2011}.} The vertical axis shows the $p$-value for a two-tailed $\chi^2$ test comparing the contingency tables obtained for the sector and for conservation (cf.~Figure~\ref{fig:dhfr_fisher}).}
	\label{fig:dhfr_secs_vs_cons_sweep}
\end{figure}

\begin{figure}
	\centering
	\includegraphics[width=0.7\textwidth]{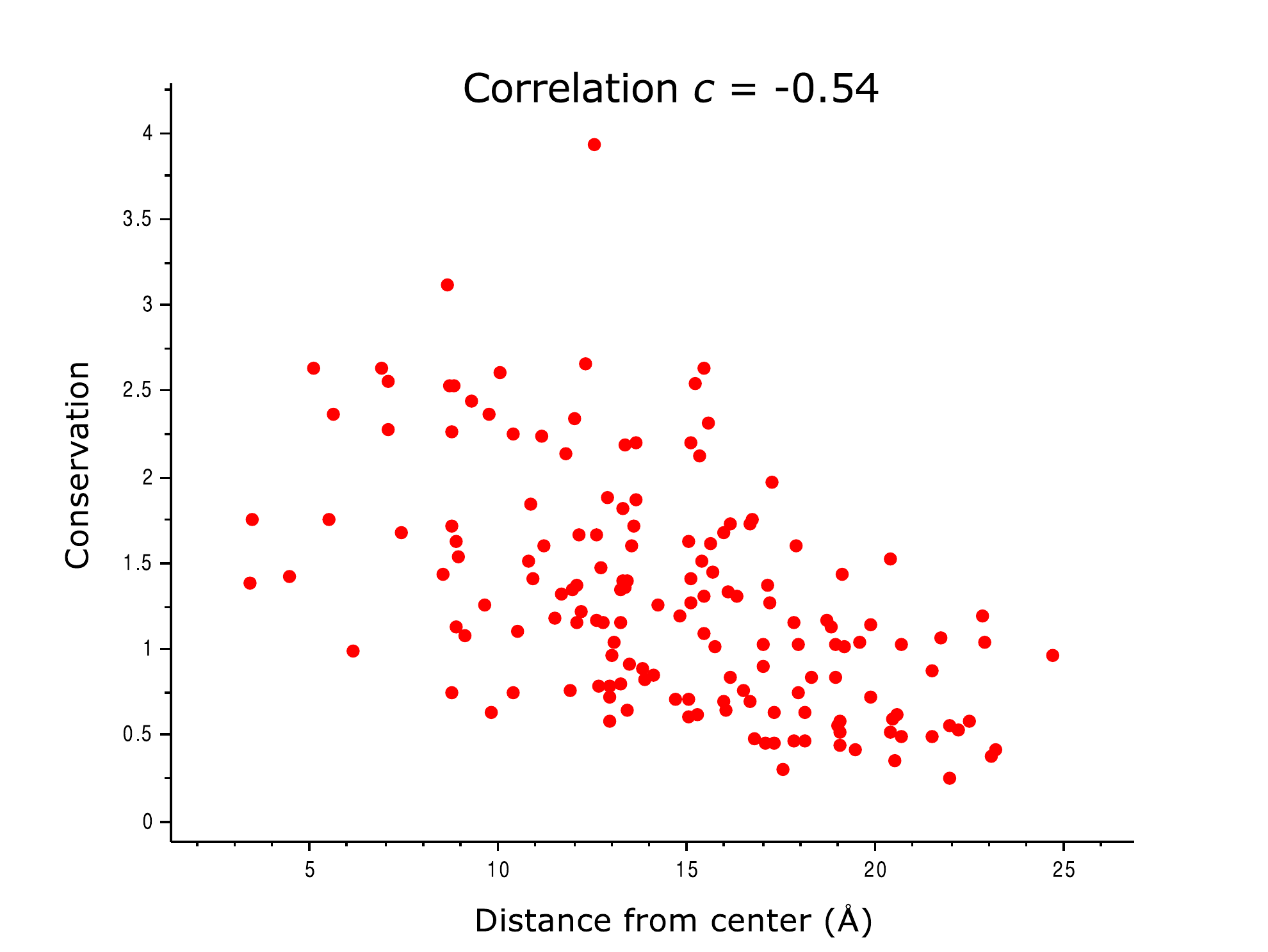}
	\caption{\textbf{Dependence of conservation level on distance from the center of mass of DHFR protein.}}
	\label{fig:dhfr_cons_vs_radius}
\end{figure}

The results we obtained for DHFR are somewhat less robust than those obtained for the other proteins. For the \hhblits DHFR alignment, the qualitative result was the same for all sector sizes we tested (see Figure~\ref{fig:dhfr_secs_vs_cons_sweep}), but when using the \pfam alignment, very small SCA sectors (less than 10 residues) ``touched'' many more functionally-significant sites than sets of conserved residues of the same size. It is hard to verify whether this is a chance occurrence or a real phenomenon, and it is unclear whether the notion of a sector still makes sense when it comprises such a small part of the protein. One complication arises from the fact that highly conserved residues tend to cluster closer to the core of the protein (see Figure~\ref{fig:dhfr_cons_vs_radius}), and thus are less likely to ``touch'' its surface.

\subsection{Voltage-sensing domains of $K^+$ channels}
Another dataset on which some work related to SCA has already been performed~\cite{Lee2009} was collected by Li-Smerin et al.~\cite{Li-Smerin2000}. In their experiments, 127 residues of the \textit{drk1} $\text{K}^+$ channel were analyzed. For each of the mutants, voltage-activation curves were measured and fit to a two-state model, from which the difference in free energy between open and closed states $\Delta G_0$ was estimated.

\begin{figure}
	\centering
	\includegraphics[width=0.6\textwidth]{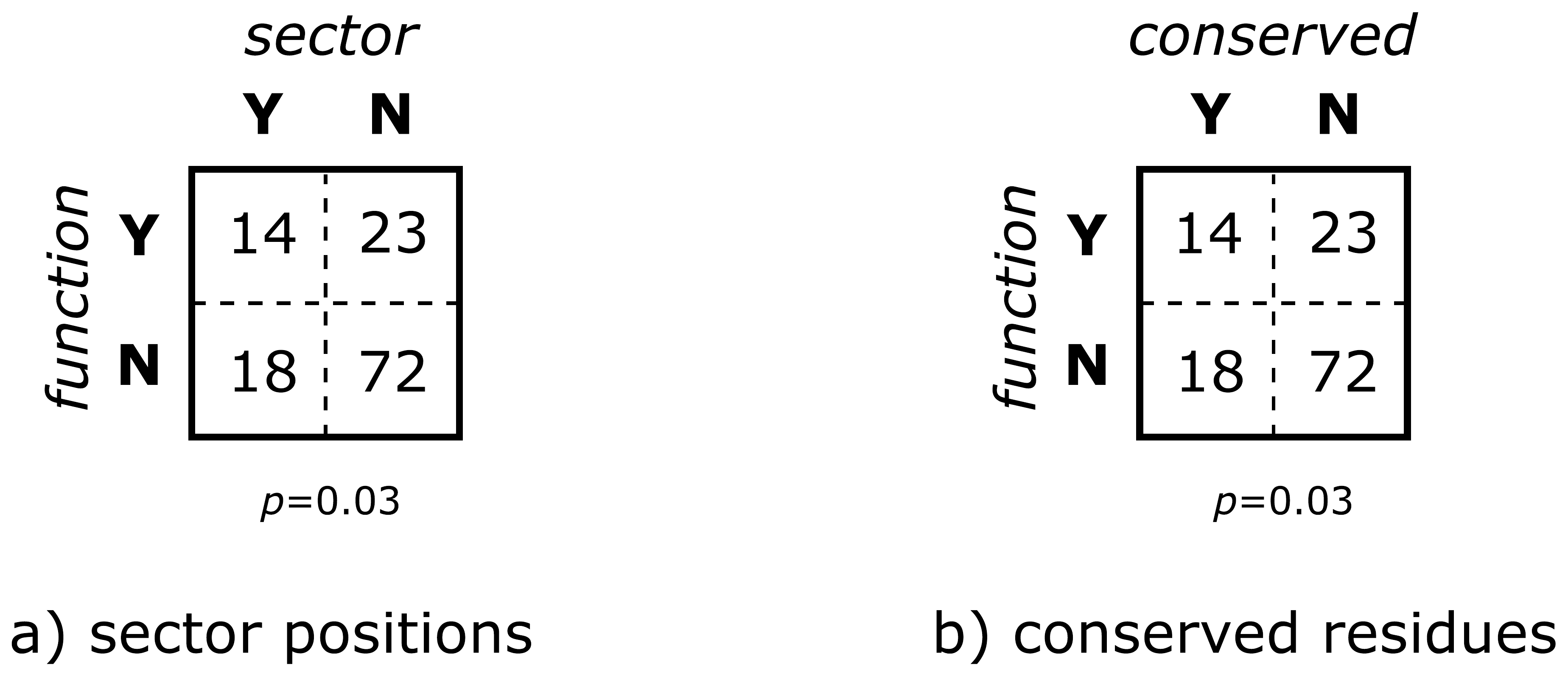}
	\caption{\textbf{Contingency tables testing whether belonging to a sector or being highly conserved is associated with significant functional effect upon mutagenesis for an alignment of voltage-sensing domains of potassium channels.} Experimental data from Li-Smerin et al.~\cite{Li-Smerin2000}. The two contingency tables are identical although less than 80\% of the residues are common between the SCA sector and the conserved positions.}
	\label{fig:kv_fisher}
\end{figure}

Following Lee et al.~\cite{Lee2009}, we identified a set of functional sites using the condition $\left\lvert\Delta G_0^{\text{mut}} - \Delta G_0^{\text{wt}}\right\rvert \ge 1\, \text{kcal/mol}$ and we compared this set to the SCA sector and to the most conserved residues. As with the other datasets, SCA and conservation turned out to be just as good at identifying functional positions in the voltage-sensing domains of potassium channels (see Figure~\ref{fig:kv_fisher}).

\subsection{\textit{E.~coli lac} repressor}

A similar dataset to the PDZ dataset described above is available for the \textit{lac} repressor protein in \text{E.~coli}~\cite{Markiewicz1994}. The authors used amber mutations and nonsense suppressor tRNAs to perform a comprehensive mutagenesis study of \textit{lacI}. In this study, each one of 328 positions was mutated to 12 or 13 alternative amino acids, and the ability of each mutant protein to repress expression of the \textit{lac} genes was tested. We summarized this data by recording, for each position, how many of the tested mutations had a significant effect on the phenotype of the \textit{lac} repressor. We further identified ``functionally-significant'' sites by considering all the positions for which at least 8 substitutions resulted in loss of function. This threshold can be varied in the whole range from 1 to 10 without significantly altering the results.

\begin{figure}
	\centering
	\includegraphics[width=0.6\textwidth]{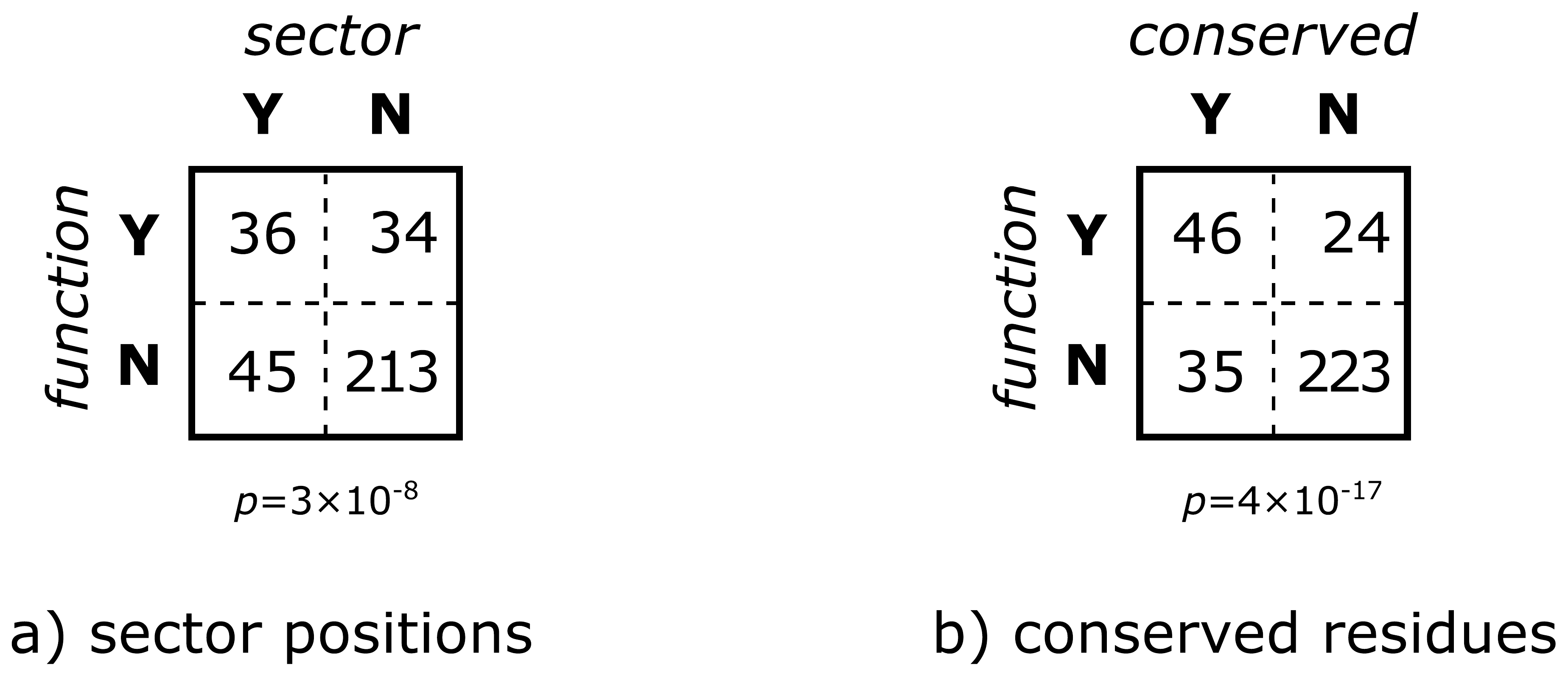}
	\caption{\textbf{Contingency tables testing whether a \textit{lac} repressor residue belonging to a sector or being highly-conserved is associated with significant functional effect upon mutagenesis.} There is about 67\% overlap between the two sets of residues. A two-tailed $\chi^2$ test cannot reject the hypothesis that the two tables are drawn from the same distribution ($p = 0.2$).}
	\label{fig:laci_fisher}
\end{figure}

\begin{figure}
	\centering
	\includegraphics[width=0.6\textwidth]{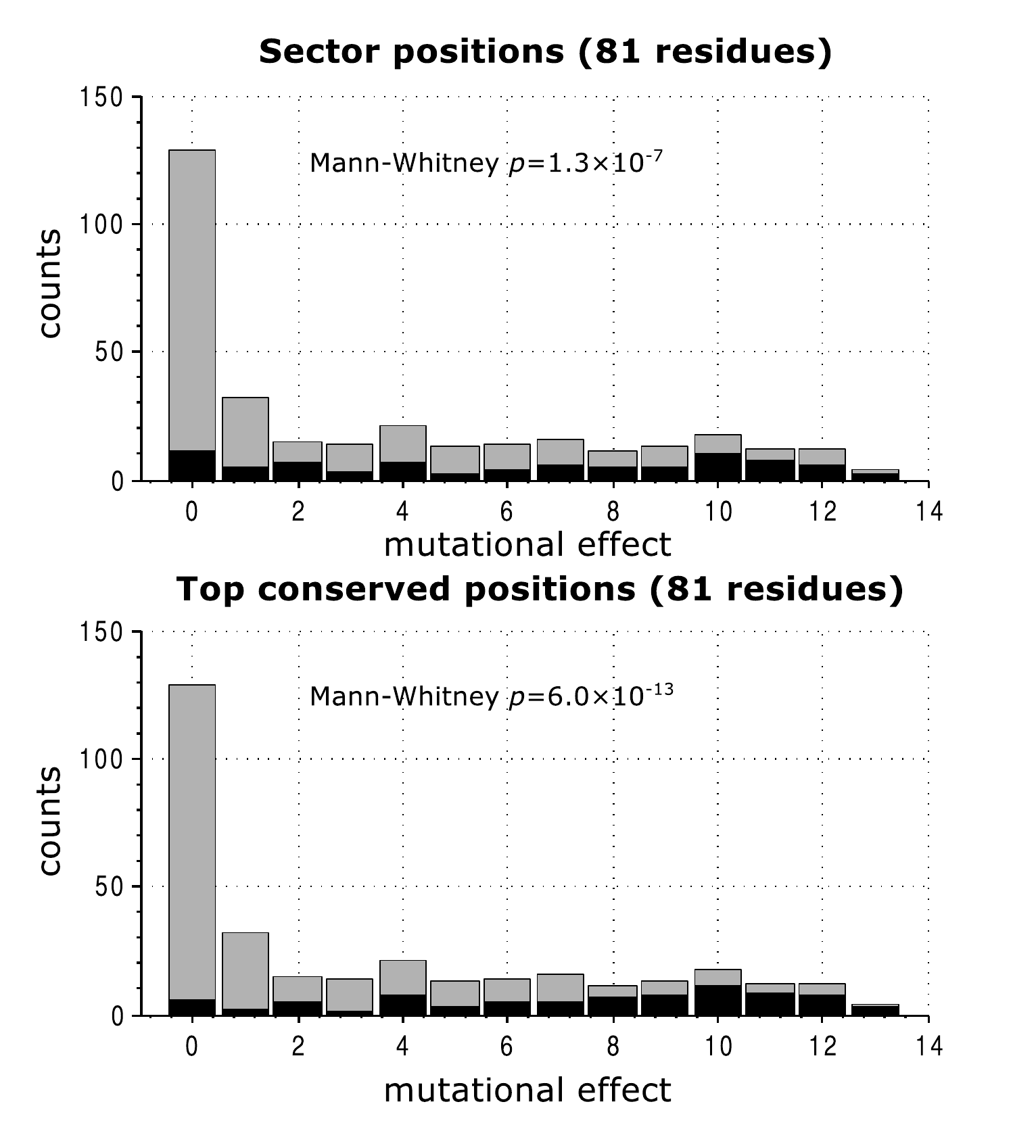}
	\caption{\textbf{Histograms showing the effect of mutations on repression ability of \textit{lacI} for all mutations (gray), and for mutations to selected positions (black).} Each of the histograms in black contains 82 positions, with the largest SCA scores (top), or the largest conservation levels (bottom). While a Mann-Whitney $U$ test finds the difference between the distribution of mutational effects for sector positions and the one for conserved positions bordering on statistical significance ($p\approx 0.08$), note that it is conservation that better matches the functional data.}
	\label{fig:laci_hist_sca_cons}
\end{figure}

\begin{figure}
	\centering
	\includegraphics[width=0.7\textwidth]{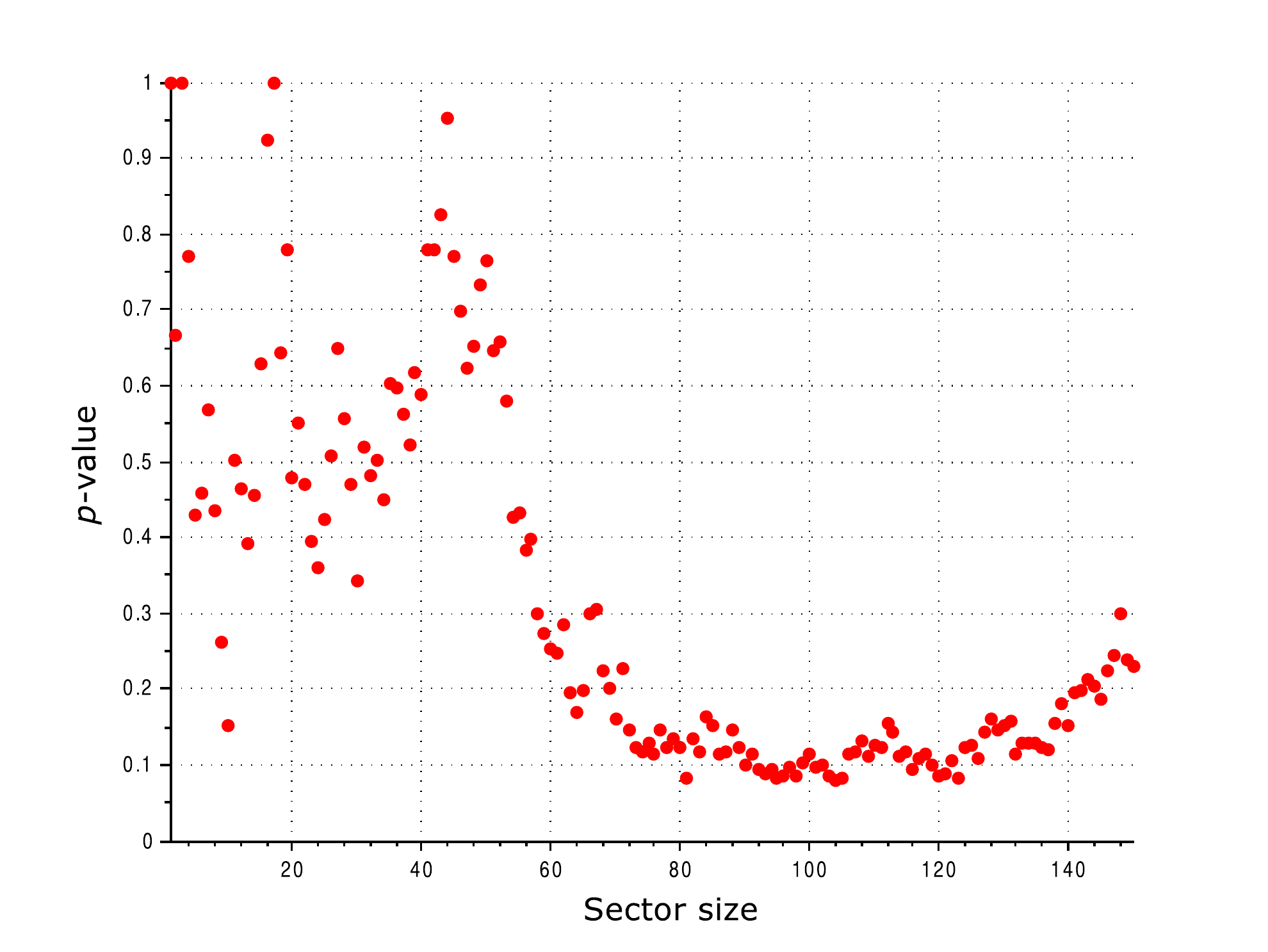}
	\caption{\textbf{Comparison of the ability of the SCA sector and conservation to predict the functional effect of mutation of \textit{lac} repressor residues for various sector sizes.} The vertical axis shows the $p$-value for a two-sample, two-tailed Mann-Whitney $U$ test comparing the distribution of mutational effects for sector residues \textit{vs.}~conserved residues. At large sector sizes, where the $p$ value hovers around $0.1$, it is the conserved residues that better match the functional data, rather than the SCA sector residues.}
	\label{fig:laci_secs_vs_cons_sweep}
\end{figure}

As before, we observed a significant association between SCA sector positions and functional positions in the \textit{lac} repressor; see Figures~\ref{fig:laci_fisher}a and~\ref{fig:laci_hist_sca_cons} (top). However, again, the set of most conserved positions was equally good at predicting functional sites---see Figures~\ref{fig:laci_fisher}b and~\ref{fig:laci_hist_sca_cons} (bottom). The results were not significantly affected by changing the size of the sector (see Figure~\ref{fig:laci_secs_vs_cons_sweep}).

\subsection{Top eigenmode of the SCA matrix}
In the previous sections, we showed that a significant fraction of the sector positions obtained from the top eigenvector of the SCA matrix can be predicted from single-site statistics. This can be attributed to a strong correlation between the components of the top eigenvector and the square root of the diagonal elements of the SCA matrix (see Figure~\ref{fig:pdz_topev_vs_diag_and_cons}, left). In Halabi et al., the top eigenvector of the SCA matrix was ignored by analogy to finance, where this mode is a consequence of global trends in the market that affect all the stocks in the same way~\cite{Halabi2009}. For proteins, the analogy is suggested to be with parts of sequences that are conserved due to phylogenetic relationships between the sequences in the alignment. Here we show that there is a different mechanism that can generate a spurious top eigenmode of the SCA matrix even when there are no phylogenetic connections between the sequences in the alignment. The main ingredient in this mechanism is a positive bias for the components of the SCA matrix.

Suppose that the underlying evolutionary process has no correlations between positions. Due to sampling noise, empirical correlations will typically be non-zero, and will fluctuate in a certain range. We denote the size of these fluctuations by $x$. The off-diagonal elements of the covariance matrix will have mean zero and variances of order $C_{ij}^2 \sim C_{ii} C_{jj} x^2$. In this case, the reason for the positive bias for the components of the SCA matrix is the fact that typically SCA takes the absolute value of the covariances (or some norm that produces only non-negative values; see appendix~\ref{sec:supp_sca})~\cite{Halabi2009, Reynolds2011, McLaughlinJr2012}. This implies that the off-diagonal entries of this matrix will have expectation values of order $x \sqrt{C_{ii} C_{jj}}$.\footnote{Note that the positional weights can be absorbed into the diagonal elements $C_{ii}$, so we do not write them out explicitly.}

Even when the absolute value is not used, the correlation between the components of the top eigenmode of the SCA matrix and the diagonal elements of this matrix may also occur; this happens for example for the alignment in Smock et al.~\cite{Smock2010}. Simulations involving random alignments show that this phenomenon occurs whenever there are weak, uniform correlations between all the positions in an alignment. This can be the result of phylogenetic bias, but could have a different origin. This situation could be distinguished from the one above by looking at how the magnitude $x$ of the off-diagonal correlations scales with alignment size; it should scale roughly like the inverse of the number of sequences if it is due to sampling noise, and be approximately constant otherwise.\footnote{We thank D.\ Hekstra for this observation.}

To try to explain these empirical observations, let us consider a simplified version of the SCA matrix:
\begin{equation}
	\label{eq:simpleScaMat0}
	M = \begin{pmatrix}
		\Delta_1 & d_1 d_2 x & \cdots & d_1 d_n x\\
		d_2 d_1 x & \Delta_2 & \cdots & d_2 d_n x\\
		\cdots & \cdots & \ddots & \cdots\\
		d_n d_1 x & d_n d_2 x & \cdots & \Delta_n
	\end{pmatrix}\,.
\end{equation}
Writing out the eigenvalue equation and performing some simple algebraic manipulations reveals that the eigenvector components $v_i$ corresponding to eigenvalue $\lambda$ are related to the diagonal elements $\Delta_i$ by
\begin{equation}
	\label{eq:evecVsFctDiag0}
	\frac {\sqrt{\Delta_i}} {v_i} \propto \lambda - \frac {\Delta_i} {1+x} \,.
\end{equation}
When the top eigenvalue is much larger than the other ones, which is usually the case when applying SCA to protein alignments, the following approximation holds:
\begin{equation}
	\label{eq:topEvalEstimate0}
	\lambda_{\text{top}} \approx \frac x {1+x} \sum_i \Delta_i\,.
\end{equation}
Empirically, this is observed to roughly match the results of SCA on real protein alignments. Given that $\lambda_{\text{top}} \gg \Delta_i$, we can also write
\begin{equation}
	\label{eq:evecVsSqrtDiag0}
	v_{i,\text{top}} \approx \frac \alpha{\lambda_{\text{top}}} \times \sqrt{\Delta_i} \,,
\end{equation}
where $\alpha$ is a normalization constant. This is the observed linear relation between the top eigenvector and the square root of the diagonal elements of the SCA matrix (Figure~\ref{fig:pdz_topev_vs_diag_and_cons}, left). Note that the SCA matrix for an alignment does not really have the highly symmetric form~\eqref{eq:simpleScaMat0}; instead it shows fluctuations in the off-diagonal components. Because of this, we cannot expect to see all the eigenvectors obey eq.~\eqref{eq:evecVsFctDiag0}. Indeed, for SCA matrices obtained from protein alignments, eq.~\eqref{eq:evecVsFctDiag0} seems to hold only for the top eigenvector. A treatment of this problem in the framework of random matrix theory might help to clear up the expectations one should have for the top eigenvector of the SCA matrix, but such an analysis goes beyond the scope of this paper.

The simple argument described above suggests that, under certain conditions that seem to hold in the cases where SCA has been applied, the top eigenvector of the SCA matrix is indeed related to conservation, and is largely independent of correlations between positions. This does not mean that there is no information contained in this top mode, but does imply that most of this information can be obtained by looking at single-site statistics alone.

Note again that in our derivation the origin of the off-diagonal entries is not specified. While we showed that they can be a simple artifact of sampling noise, they could also be partly due to a non-trivial phylogenetic structure of the alignment, as previously suggested~\cite{Halabi2009}.

\section{Discussion}

\subsection{Proteins with multiple SCA sectors}
It is perhaps not surprising that conservation is a good indicator of the functionally-important residues in a protein; indeed, this fact is one of the original motivations for using positional weights in SCA that grow with conservation levels~\cite{Ranganathan2011url}. However, as a consequence, for proteins with a single SCA sector, it is difficult to distinguish between the functional significance of sector residues and that of conserved residues. The natural solution to this problem is to focus on proteins with multiple sectors, such as the serine protease family analyzed by Halabi et al.~\cite{Halabi2009}.

In the serine protease case, three SCA sectors were identified by placing thresholds on certain linear combinations of eigenvectors of the SCA matrix.\footnote{The top eigenvector was ignored based on an analogy to finance, and thus the issues outlined in the previous section do not apply here.} These sectors (called `blue', `red', and `green') were found to have independent effects on various phenotypes of the protein: the blue sector affected denaturation temperature, the red one affected binding affinity, and the green sector contained the residues responsible for catalytic activity.

There are two attractive features of the serine protease data. One is that several different quantities were measured for each mutant, thus allowing for a test of the idea that the protein is split into groups each of which affects different phenotypes. Another important feature is that some double mutants were also measured, showing that mutations in different sectors act approximately independently from each other. Collecting more extensive data of this type for serine proteases and for other proteins should give more weight to the idea that SCA sectors act as functional sectors in proteins. To reduce the amount of work involved, we point out that from our observations, it seems that instead of a complete scan of all 19 alternative amino acids at each position, an alanine scan, involving only mutations to alanine, might be sufficient. Using only alanine replacements, even a complete double-mutant study of \psd~would require about 3000 mutants, only a factor of two more than were already studied~\cite{McLaughlinJr2012}. For proteins exhibiting multiple SCA sectors, this number could be lowered by focusing only on those double mutants that combine mutations in different sectors, thus testing the independence property.

Finding several relevant quantities to measure for each of the mutants might not be an easy task. An ideal system for this would be related to gene expression or signal transduction, allowing measurements to be made in realistic conditions. Furthermore, it would be convenient to have a low-dimensional quantitative description of the protein's phenotype, so that one could check whether the sectors predicted by SCA correlate with the mutations that affect the parameters in this description.

One difficulty in the application of SCA is that the identification of sectors is non-trivial. Halabi et al.\ used visual inspection to identify linear combinations of eigenvectors to represent the sectors~\cite{Halabi2009}. Independent component analysis (ICA) has also been invoked to find the linear combinations~\cite{Smock2010, Ranganathan2011url, Rivoire2012}, but a mathematically rigorous motivation for the application of this procedure is missing. An approach that avoids these difficulties is to check whether a linear regression can approximate the measured quantities for the different mutants with linear combinations of the eigenvectors of the SCA matrix. This seems to work for the case of serine protease (see Figure~\ref{fig:supp_lincomb_trypsin}), though the small number of data points prevents a statistically rigorous analysis. A similar approach does not work for the PDZ data from McLaughlin Jr.\ et al., in which binding to both the cognate (CRIPT) ligand and to a mutated T$_{-2}$F ligand was measured~\cite{McLaughlinJr2012} (see Figure~\ref{fig:supp_lincomb_pdz}). It also does not work for the potassium channels dataset, in which both the activation voltage $V_{50}$ and the equivalent charge $z$ were measured for each mutant~\cite{Li-Smerin2000} (see Figure~\ref{fig:supp_lincomb_kv}). This is consistent with the idea that these proteins exhibit a single sector.

Conservation alone cannot in general be used to find several distinct groups of residues that have distinct functions. For this reason, finding evidence for functionally significant and independent SCA sectors would automatically favor SCA over a simple conservation analysis. However, it is important to point out that SCA, with the particular set of weights as defined by Halabi et al.~\cite{Halabi2009}, is only one possible procedure for analyzing correlations in sequence alignments. Once more data is available for proteins containing multiple sectors, it will be important to compare different sets of positional weights, or different models altogether, to identify the best approach for analyzing MSAs~\cite{Colwell2014}.

\subsection{Conclusions}
We analyzed the available evidence regarding the hypothesis that the residues comprising the sectors identified by statistical coupling analysis are functionally significant. We looked at a number of studies, some directly related to SCA~\cite{Halabi2009, Reynolds2011, McLaughlinJr2012}, and some unrelated~\cite{Li-Smerin2000, Markiewicz1994}, and we showed that while the sector positions identified by SCA do tend to be functionally relevant, in the case of single-sector proteins, conserved positions provide a statistically equivalent match to the experimental data. This observation was traced to a property of the SCA matrix that makes the components of its top eigenvector correlate strongly with its diagonal entries. We presented a mathematical model that might explain this correlation. This model suggests that, as a generic property of statistical coupling analysis, the top eigenvector of the SCA matrix does not contain information beyond that provided by single-site statistics.

The observation that conservation is an important determinant of the SCA sectors is of course not very surprising, since one of the principles of SCA is to upweight the correlation information for conserved residues compared to poorly-conserved ones. However, this does pose a problem for the interpretation of the large-scale experiments that have been performed in relation to SCA~\cite{Reynolds2011, McLaughlinJr2012}, given that these provide most of the available evidence for the functional significance of SCA sectors. Our analysis show that this functional significance might be due to conservation alone. Since function is not the only reason for which protein residues may be conserved~\cite{Mirny1999}, it is not surprising that the overlap with functional residues is not perfect.

Once again, it is important to note that our findings do not imply that correlations within MSAs are uninformative; the contrary seems to be supported by experimental data~\cite{Russ2005, Weigt2009, Morcos2011, Marks}. However, in order to test whether the particular way in which these correlations are used within the SCA framework is useful for making functional predictions about proteins, it will be necessary to go beyond single-sector proteins and measure several different phenotypes. Such data exists~\cite{Halabi2009}, but is too limited at this point to be conclusive. A thorough investigation of the idea that SCA sectors act as functional sectors requires more of this type of data, for a wider class of proteins.

Whether small groups of residues inside proteins act as independent ``knobs'' controlling the various phenotypes is a question that can be asked independently of any statistical analysis of alignments. Such functional sectors could be found by mutagenesis work, as described above. Alternatively, one could look for structural sectors using NMR or X-ray data to search for correlated motions. This has the advantage of not requiring the modification of proteins through mutations. Finally, evolutionary sectors could be searched for by using artificial evolution experiments. If the existence of these functional, structural, or evolutionary sectors is verified with sufficient precision, one could then more easily approach the question of whether a statistical method is capable of inferring their composition from an MSA, and in this case, which method is the most efficient and accurate.

\section{Methods}

\subsection{Sequence alignments}
\label{sec:methods_alignments}
Statistical coupling analysis requires an alignment of protein sequence homologs as input data. This may contain both orthologs and paralogs, and at least moderate sequence diversity within the alignment is necessary, because an alignment of identical sequences will not contain any information about amino acid covariance. The alignments we used were generated using \hhblits, with an $E$-value of $E = 10^{-10}$. States with 40\% or more gaps were considered insert states, and were later removed from the calculations. The Uniprot IDs of the seed sequences used with \hhblits are as follows: DLG4\_RAT (PDZ), DYR\_ECOLI (DHFR), KCNB1\_RAT ($K^+$ channels), and LACI\_ECOLI (\textit{lacI}). To check the robustness of the results, we also ran our analysis on \pfam alignments when available, and on the alignments from McLaughlin Jr.\ et al.~\cite{McLaughlinJr2012}, Reynolds et al.~\cite{Reynolds2011}, and from Lee et al.~\cite{Lee2009} for the PDZ, DHFR, and potassium channels datasets, respectively.

\subsection{Statistical coupling analysis}
\label{sec:methods_sca}
The statistical coupling analysis was performed in accordance with the projection method~\cite{McLaughlinJr2012, Ranganathan2011url}, which is the default in the newest version of the SCA framework from the Ranganathan lab. The code we used for the analysis can be accessed at \textbf{https://bitbucket.org/ttesileanu/multicov}.

Consider a multiple sequence alignment represented as an $N \times n$ matrix $A$ in which $a_{ki}$ is the amino acid at position $i$ in the $k^{\text{th}}$ sequence. We first construct a numeric matrix $\tilde X$ by
\begin{equation}
	\label{eq:projectionMethodDesc}
	\tilde x_{ki} = \begin{cases}
		\cfrac {\phi_i(a_{ki}) f_i(a_{ki})} {\sqrt{\displaystyle\sum\limits_{b\ne \text{gap}} \phi_i^2(b) f_i^2(b)}} & \text{if $a_{ki} \ne \text{gap}$}\,,\smallskip\\
		0 & \text{if $a_{ki} = \text{gap}$}\,,
	\end{cases}
\end{equation}
where $\phi_i(a)$ is a positional weight, and $f_i(a)$ the frequency with which amino acid $a$ occurs in column $i$ of the alignment. The positional weights are given by
\begin{equation}
	\label{eq:positionalWeights}
	\phi_i(a) = \log \left[\frac {f_i(a)} {1 - f_i(a)} \frac {1 - q(a)} {q(a)}\right]\,,
\end{equation}
where $q(a)$ is the background frequency with which amino acid $a$ occurs in a large protein database. The SCA matrix is, up to an absolute value, the covariance matrix associated with $\tilde X$,
\begin{equation}
	\tilde C_{ij} = \left\lvert \frac 1N\sum_k \tilde x_{ki} \tilde x_{kj} - \frac 1 {N^2}\sum_{k, l} \tilde x_{ki} \tilde x_{lj}\right\rvert\,.
\end{equation}
Finally, the sector was identified by finding the positions where the components of the top eigenvector of $\tilde C_{ij}$ went above a given threshold. The threshold was chosen so that the sector comprised about 25\% of the number $n$ of residues contained in each alignment sequence.

More details about this method and descriptions of the other variants of SCA found in the literature can be found in appendix~\ref{sec:supp_sca}.

\subsection{Sequence conservation}
The conservation level of a position in the alignment is calculated using the relative entropy (Kullback-Leibler divergence), as described in eq.~\eqref{eq:defConservation0}. A different definition, as the frequency of the most prevalent amino acid at a position, is highly correlated with $D_i$ and gives similar results.

Note that the calculation of the relative entropy as defined in eq.~\eqref{eq:defConservation0} requires that $\sum_a f_i(a) = 1$ and $\sum_a q(a) = 1$. For the first of these relations to hold, we need the sum over $a$ to include the gap, but this requires a value for the background frequency of gaps $q(\text{gap})$. This is not straightforward to estimate or even to define. There are several solutions possible: one is to assume that the background frequency for gaps is equal to the gap frequency in the alignment averaged over all positions. Another approach is to simply ignore the gaps by focusing only on the sequences that do not contain a gap at position $i$. We chose the former solution, as it is the default one in the SCA framework, but the results are very similar when using the latter choice.

\section{Acknowledgments}
We are grateful to Richard McLaughlin Jr., Rama Ranganathan, and Kim Reynolds for sharing their scripts and data with us, and for useful discussions. We would also like to thank G\'erard Ben Arous, Doeke Hekstra, Michael Mitchell, Rama Ranganathan, and Olivier Rivoire for discussions and comments on early drafts of this manuscript.  T.\ T.\ was supported by a Charles L.\ Brown Membership at the Institute for Advanced Study. L.\ J.\ C.\ was supported by an Engineering and Physical Sciences Research Council Fellowship (EP/H028064/2).

\newcommand{\compactfisher}[5]{
	\begin{tabular}{rccc}
			&		\textbf{#5}		&		\textbf{N#5}		&	\\
			\cmidrule(lr){2-3}
			\textbf{F}	&		#1		&		#2		&	\\
			\textbf{NF}	&		#3		&		#4 	&
	\end{tabular}
}

\newpage
\appendix

\renewcommand{\thefigure}{S\arabic{figure}}
\setcounter{figure}{0}
\renewcommand{\theequation}{S\arabic{equation}}
\setcounter{equation}{0}
\renewcommand{\thetable}{S\arabic{table}}
\setcounter{table}{0}

\section{Calculating sequence covariance}
\label{sec:covmats}
	
Let us assume we are given a multiple sequence alignment of homologous proteins as an $N\times n$ matrix $A_{ki}$. In order to calculate correlations in the alignment, we need a way of transforming it to numeric data. There are three main ways in which this was done in the literature~\cite{Halabi2009, Smock2010, Reynolds2011, McLaughlinJr2012, Rivoire2012}, which we describe below; generally all these methods yield highly similar results. In the present work, we used the projection method (item~\ref{it:projmeth} below), which is the default in the latest SCA release from the Ranganathan lab. Note that SCA requires positional weighting to be done on top of the covariance analysis. This is described in the next section.

\begin{enumerate}
	\item \textbf{The binary approximation.}
		
	We start by describing the simplest approach, which constructs a ``binary approximation'' of the alignment~\cite{Halabi2009, Smock2010}. In the binary approximation, each amino acid is replaced by 1 if it is equal to the consensus amino acid at its position, and 0 otherwise,
	\es{eq:defBinApprox}{X_{ki} = \delta(A_{ki}, c_i)\,,}
	where the consensus amino acid $c_i$ is the most frequent amino acid found in column $i$ of the alignment.\footnote{This is typically restricted to non-gaps, though in practice this usually does not affect the results.} Here $A_{ki}$ is the amino acid found at position $i$ in sequence $k$, and the Kronecker symbol $\delta(a, b)$ is 1 if and only if $a=b$.
		
	The covariance matrix is then defined in the standard way,
	\es{eq:defCov}{C^{\text{bin}}_{ij} = f_{ij} - f_i f_j \,,}
	where
	\es{eq:defFreqs}{f_{ij} = \frac 1N \sum_k X_{ki} X_{kj} \,, \qquad f_i = \frac 1N \sum_k X_{ki}\,,}
	with $N$ being the number of sequences in the alignment. Note that $f_i$ is simply the frequency of the consensus amino acid at position $i$, which is often used as a measure of the conservation level at that position. Also, $f_{ij}$ is the frequency at which the consensus amino acids occur simultaneously in the two columns $i$ and $j$.
		
	\item \textbf{The reduction method.}
	\label{it:redmeth}
		
	The most generic statistical analysis that can be performed with categorical data is using \emph{contingency tables}. In this context, these are tables of the frequencies at which various combinations of amino acids occur simultaneously in a sequence---we can, for example, define the frequency $f_i(a)$ with which amino acid $a$ is found at position $i$, and the frequency $f_{ij}(a, b)$ with which amino acids $a$ and $b$ co-occur at positions $i$ and $j$, respectively. It is then convenient to define a ``binary representation'' of the alignment~\cite{Rivoire2012}, $x_{ki}(a)$, where $x_{ki}(a)$ is equal to 1 if $A_{ki}$ is $a$, and 0 otherwise; in short,
	\es{eq:defBinRep}{x_{ki}(a) = \delta(A_{ki}, a)\,.}
	Note that here there are 21 columns for each column in the original alignment. It is important to keep in mind that---despite the potentially confusing nomenclature---the binary representation $x_{ki}(a)$ is very different from the binary approximation $X_{ki}$. The former is an exact representation of the alignment data, while the latter is an approximation that keeps only part of the information in $A_{ki}$.
		
	The single-site and pairwise frequencies, $f_i(a)$ and $f_{ij}(a, b)$, are thus averages involving the binary representation
	\es{eq:defFreqsFull}{
		f_i(a) &= \frac 1N \sum_k x_{ki}(a) \,,\\
		f_{ij}(a, b) &= \frac 1N \sum_k x_{ki}(a) x_{kj} (b)\,.
	}
	Now we can define the covariance
	\es{eq:defCovFull}{C_{ij}(a, b) = f_{ij}(a, b) - f_i(a) f_j(b)\,.}
	For each pair of sites, we obtain a covariance value between every pair of amino acids, $a$ and $b$. There are 21 choices of amino acid (including the gap), so for each pair of sites $(i, j)$, $C_{ij}$ is a $21\times 21$ matrix. Note, however, that frequencies are normalized---$\sum_a f_i(a) = 1$ and $\sum_b f_{ij}(a, b) = f_i(a)$---which means that not all numbers in the covariance matrix are independent. In fact, if we are given the values of $C_{ij}(a, b)$ for all amino acids except one (commonly, the gap), it is straightforward to infer the missing values. For this reason, unless otherwise stated, we will always assume $a$ and $b$ to exclude the gap, and will treat the matrix $C_{ij}$ as having size $20\times 20$ instead of $21\times 21$.
		
	In the binary approximation, for each pair of sites $i$ and $j$, we were able to calculate one number $C^{\text{bin}}_{ij}$ showing the amount of covariance between the sites. In contrast, the full covariance matrix $C_{ij}(a, b)$ contains an entire matrix of numbers for each pair of sites. It is sometimes useful to collapse this matrix to a single number that measures the overall covariance between two sites, as was the case in the binary approximation. This is usually done using a heuristic approach, for example by using the Frobenius norm~\cite{Ranganathan2011url}
	\es{eq:frobRed}{C_{ij}^{\text{red}} = \biggl[\sum_{a,b} C_{ij}^2(a,b)\biggr]^{1/2}\,,}
	or the spectral norm~\cite{Reynolds2011}
	\es{eq:specRed}{C_{ij}^{\text{red}} = \text{largest singular value of $C_{ij}(a, b)$}\,.}
	The binary approximation described above can actually also be seen as a reduction method, in which
	\es{eq:binRed}{C_{ij}^{\text{red}} = C_{ij}(c_i, c_j)\,,}
	with $c_i$ being the consensus amino acid in column $i$, as before.
		
	Another approach for carrying out the reduction starts from a ratio of frequencies,
	\es{eq:defPreMI}{D_{ij}(a, b) = \log \frac {f_{ij}(a, b)} {f_i(a) f_j(b)} = \log \biggl[1 + \frac {C_{ij}(a, b)} {f_i(a) f_j(b)}\biggr]\,.}
	The mutual information is then a natural measure of the independence of two random variables that can be constructed from $D$ \cite{Fodor2004, Dunn2008},
	\es{eq:defMI}{\MI_{ij} = \!\!\sum_{\substack{a, b\\\text{incl. gaps}}} \!\!\!\! f_{ij}(a, b) D_{ij}(a, b) = \!\!\sum_{\substack{a, b\\\text{incl. gaps}}} \!\!\!\! f_{ij}(a, b) \log \frac {f_{ij}(a, b)} {f_i(a) f_j(b)}\,.}
	
	Note that here we include gaps in the summation because the definition of the mutual information given in eqs.~\eqref{eq:defPreMI} and~\eqref{eq:defMI} assumes that $f_i(a)$ and $f_{ij}(a, b)$ are proper frequencies; in particular, they should sum up to 1, and this is only true if the values for the gaps are included in the sum.
		
	\item \textbf{The projection method.}
	\label{it:projmeth}
	\nopagebreak
		
	Instead of calculating the full covariance matrix and then reducing it, the projection method starts by projecting the binary representation onto an $N\times n$ numeric matrix $Y_{ki}$,
	\es{eq:defProjApprox}{
		Y_{ki} := \sum_a x_{ki}(a) v_i(a) \,,
	}
	where $v_i$ are unit vectors. We can then use the definitions~\eqref{eq:defCov} and~\eqref{eq:defFreqs} with $Y$ instead of $X$ to obtain the covariance matrix,
	\es{eq:defCovProj}{
		C_{ij}^{\text{proj}} = \frac 1N \sum_k Y_{ki} Y_{kj} - \frac 1 {N^2} \sum_{k,l} Y_{ki} Y_{lj}\,.
	}
	The projection vectors that we use in this paper are given by~\cite{McLaughlinJr2012}:
	\es{eq:ramaProj}{
		v_i(a) = \frac {f_i(a)} {\sqrt{\sum_b f_i(b)^2}}\,.
	}
	Note that, as we did above for the covariance matrix, here too we do not include the gap, \textit{i.e.,} we assume that $v_i(\text{gap}) = 0$, and we have $b \ne \text{gap}$ in the summation.
	
	It is interesting to note that a different choice for the $v_i$ can be used to recover the binary approximation,
	\es{eq:binAsProj}{
		v_i^{\text{binary}}(a) = \delta\bigl(a, \argmax_{b \ne \text{gap}} f_i(b)\bigr) \equiv \delta(a, c_i)\,.
	}
	For both the binary approximation and the projection method, the approximation is most accurate for highly conserved sites.
\end{enumerate}

\section{Statistical coupling analysis}
\label{sec:supp_sca}
There are two main features that distinguish SCA from a standard covariance analysis: the introduction of positional weights and the positivity of the matrix elements. Starting again with the binary approximation, for simplicity, the covariance matrix $C$ defined in~\eqref{eq:defCov} is transformed to~\cite{Halabi2009}
\es{eq:scaBinMatrix}{\tilde C_{ij} = \abs{\phi_i \phi_j C_{ij}}\,,}
where the weights $\phi_i$ are specific functions of the single-site statistics, chosen based on the idea that entries of the covariance matrix corresponding to poorly conserved sites are less likely to be informative than those corresponding to highly conserved sites, and thus should be given less weight.
	
The absolute value that appears in the formula was justified by the requirement of finding blocks of coevolving residues regardless of the sign of the correlations~\cite{Halabi2009}. It also avoids a certain instability to small perturbations that appears in cases in which the consensus amino acid has a frequency that is very similar to that of the next most common amino acid. In these cases, a small perturbation in the alignment can flip the order of the top amino acids, thus flipping the sign of some correlations in the binary approximation.\footnote{We thank O.~Rivoire for this observation.} Taking the absolute value may, however, introduce artifacts, as described in the main text.
	
The expression commonly used for the positional weights is~\cite{Halabi2009, Smock2010, McLaughlinJr2012}
\es{eq:stdPosWeights}{\phi_i = \log \left[\frac {f_i}{1 - f_i} \frac {1 - q(c_i)} {q(c_i)}\right]\,,}
where $f_i$ is the conservation at site $i$, \textit{i.e.}~the frequency in the alignment of the consensus amino acid, while $q(c_i)$ are background frequencies for the consensus amino acids $c_i$. Note that we are assuming that the background frequencies depend only on the identity of the consensus amino acid, and not on the position $i$ in the protein. These background expectations can be estimated by averaging over a large set of proteins. The functional form~\eqref{eq:stdPosWeights} for the positional weights was chosen to match the original 1999 formulation of SCA~\cite{Lockless1999, Halabi2009} and to fulfill the role of down-weighting poorly conserved sites, but is otherwise arbitrary.
	
Instead of performing the positional weighting at the level of the covariance matrix, it could have been performed on the binary alignment itself. The covariance matrix of the transformed binary alignment
\es{eq:weightedBin}{\tilde X_{ki} = \phi_i X_{ki}}
directly yields $\tilde C_{ij}$, after taking the absolute value of each element. This can be generalized to apply to the binary representation matrix; we define
\es{eq:weightedBinRep}{\tilde x_{ki}(a) = \phi_i(a) x_{ki}(a)\,,}
in which we replace~\eqref{eq:stdPosWeights} by~\cite{Rivoire2012}
\es{eq:stdWeightsFull}{\phi_i(a) = \log \left[\frac {f_i(a)} {1-f_i(a)} \frac {1 - q(a)} {q(a)}\right]\,.}
With this positionally-weighted binary representation we can use either the reduction method~\cite{Rivoire2012} or the projection method~\cite{McLaughlinJr2012} described in the previous section. When using the projection method, the absolute value of each element is taken when calculating the SCA matrix, as is done with the binary approximation. An interesting empirical observation is that the mutual information defined in eq.~\eqref{eq:defMI} is well-approximated by a weighted SCA matrix using the reduction method, in which the positional weights are chosen equal to the logarithm of the frequencies.\smallskip

For the results in the main text, after the SCA matrix was calculated, the sector positions were identified by placing a threshold on the components of its top eigenvector. Note that, due to the positivity of the elements of the SCA matrix, the Perron-Frobenius theorem guarantees that all these components are positive.

\section{Alignments}
\label{sec:supp_alignments}
As described in the Methods section of the paper, the software package \hhblits~\cite{Remmert2012} was used to generate the alignments used in the the main text. We also analyzed a number of alignments from other sources to make sure the results weren't sensitive to this choice. For the cases of PDZ~\cite{McLaughlinJr2012}, DHFR~\cite{Reynolds2011}, and the potassium channels~\cite{Lee2009}, we ran the analysis on the alignments that were used in the papers that first applied the SCA method to those proteins. For PDZ and DHFR we also ran the analysis on \pfam alignments (\texttt{PF00595} and \texttt{PF00186}, respectively). This was not done for the potassium channels and \textit{lacI} because no suitable alignments were available in \pfam. See Table~\ref{tab:analysis_on_old_alignments} for a summary of the results.

\begin{table}
	\caption{Results of the analysis performed in this paper when run on different alignments.\label{tab:analysis_on_old_alignments}}
	\bigskip
	\begin{minipage}{\linewidth}
	\centering
	\begin{tabular}{>{\centering}p{0.17\textwidth}>{\centering}p{0.25\textwidth}>{\centering}p{0.25\textwidth}>{\centering}p{0.17\textwidth}}
		Alignment & Contingency table for sector\footnote{\textbf{C}, \textbf{S}, and \textbf{F} stand for conserved, sector, and functional, respectively.} & Contingency table for conservation & Comparison ($\chi^2$ $p$ value)
		\bigskip\tabularnewline
		\toprule\vspace{3em}\tabularnewline
		PDZ~\cite{McLaughlinJr2012} & \compactfisher{15}{5}{8}{53}{S} & \compactfisher{15}{5}{8}{53}{C} & $p_{\chi^2} = 1.00$\tabularnewline[3em]
		PDZ (\pfam) & \compactfisher{13}{7}{7}{52}{S} & \compactfisher{9}{11}{11}{48}{C} & $p_{\chi^2} = 0.45$\tabularnewline[3em]
		DHFR\footnote{For DHFR we are counting the residues that are ``touched'' by either sector or conserved residues. Although the number of conserved residues we are considering is equal to the number of residues in the sector, the number of surface residues that are ``touched'' is different.\label{tabfoot:dhfr_touch}}~\cite{Reynolds2011} & \compactfisher{14}{0}{33}{17}{S} & \compactfisher{12}{2}{27}{23}{C} & $p_{\chi^2} = 0.30$\tabularnewline[3em]
		DHFR\footnote{See footnote~\ref{tabfoot:dhfr_touch}.} (\pfam) & \compactfisher{14}{0}{35}{15}{S} & \compactfisher{12}{2}{29}{21}{C} & $p_{\chi^2} = 0.29$\tabularnewline[3em]
		potassium channels\footnote{The number of residues in the SCA sector is the same as the number of conserved residues we considered. The sum of the entries in the \textbf{S} column does not match that in the \textbf{C} column because some of the sector residues are located at sites where we have no experimental data.}~\cite{Lee2009} & \compactfisher{18}{19}{17}{67}{S} & \compactfisher{18}{19}{20}{64}{C} & $p_{\chi^2} = 0.96$
	\end{tabular}
	\end{minipage}
\end{table}

To improve the quality of the alignments, it has been suggested~\cite{Rivoire2012} to filter them by removing repeated sequences, sequences with too many gaps, and positions with too many gaps. From our tests, however, these procedures do not make a big difference to the results of this paper. For this reason, the only filtering we perform is to remove the insert states from \hhblits alignments, which in our case amounts to removing columns containing 40\% or more gaps.

In addition, for the PDZ alignment from McLaughlin Jr.~et al.~\cite{McLaughlinJr2012}, we removed the columns containing more than 20\% gaps, because this is how that alignment was processed in the scripts provided by the authors of that article~\cite{McLaughlinJr2012}. There is a minor glitch in the procedure of mapping alignment columns to PDB coordinates in McLaughlin Jr.~et al.\ that leads to the misidentification of one of the columns (corresponding to PDB position 334). For consistency with the older work, we worked with this minor error in the alignment, but we checked that the results are not significantly affected by it.

\section{Details about the DHFR analysis}
\label{sec:dhfrdetails}
	
The way in which the DHFR alignment was analyzed by Reynolds et al.~\cite{Reynolds2011} has a number of peculiarities compared to the other datasets we presented, which we describe below. We note, however, that their results are not significantly different from those obtained with our simplified protocol.

The SCA method used by Reynolds et al.\ was the spectral-norm reduction method described above (see section~\ref{sec:covmats}, item~\ref{it:redmeth}) using a thresholded form of the positional weights~\cite{Reynolds2011},
\es{eq:weightsDhfr}{\phi_i(a) = \begin{cases}
	\displaystyle\log \left[\frac {f_i(a)} {1-f_i(a)} \frac {1 - q_a} {q_a}\right] & \text{for $f_i(a) > q_a$,}\\
	0 & \text{else.}
	\end{cases}
}
Furthermore, the SCA matrix was ``cleaned'' by subtracting the average SCA matrix calculated for 100 randomized alignments. Each of the randomized alignments was obtained by independently permuting the elements of the alignment columns, which has the effect of destroying correlations without affecting the single-site amino acid frequencies.\footnote{Since the number of random samples is finite, the results depend slightly on the state of the random number generator. The results from the original paper by Reynolds et al.\ can be obtained by using the default random number generator in Matlab with the default seed.}
	
The sector was defined by the residues for which the component of at least one of the top five eigenvectors goes above a given threshold~\cite{Reynolds2011}. To select the threshold, first a Student's $t$-distribution was fit to the components of each of the eigenvectors, and then the value for which the $t$-distribution PDF drops below a certain threshold was used as a cutoff. The PDF threshold is given by $p_i$ for the $i^{\text{th}}$ eigenvector, where $p_i$ is\footnote{The dependence of the threshold on $i$ is an artifact of the fact that Reynolds et al.\ applied a constant threshold to histogram values instead of PDF values, and the bin size for the histograms was determined using the Freedman-Diaconis rule~\cite{Freedman1981}, and thus varied between eigenvectors. This can be seen from the Matlab scripts provided by the authors.}
\es{eq:dhfrThresholds}{p_i = \frac {0.005} {2 \iqr(v_i) n^{-1/3}}\,,}
where $\iqr(v)$ is the interquartile range of $v$ (the range over which the middle half of the components of $v$ spread), and $v_i$ is the $i^\text{th}$ eigenvector. Despite the complicated selection procedure, a very similar sector can be obtained by using a constant PDF threshold for each of the top five eigenvectors, or even just by using the largest components of only the top eigenvector.

\section{Alanine scans}
Instead of using all the experimental data for PDZ and for \textit{lacI}, we can restrict our attention to alanine mutations, to get an idea for the amount of information contained in an alanine scan. As mentioned in the paper, the qualitative results do not change much, though, as expected, the quality of the match between the predictions from SCA or conservation and the experimental data is reduced (see Table~\ref{fig:supp_pdz_laci_ala}).
	
\begin{table}
	\centering
	\caption{The results from the paper when restricting to alanine mutations.\label{fig:supp_pdz_laci_ala}}
	\bigskip
	\begin{tabular}{>{\centering}p{0.17\textwidth}>{\centering}p{0.25\textwidth}>{\centering}p{0.25\textwidth}>{\centering}p{0.17\textwidth}}
		Alignment & Contingency table for sector & Contingency table for conservation & Comparison ($\chi^2$ $p$ value)
		\bigskip\tabularnewline
		\toprule\vspace{3em}\tabularnewline
		PDZ & \compactfisher{10}{10}{11}{50}{S} & \compactfisher{9}{11}{12}{49}{C} & $p_{\chi^2} = 0.98$\tabularnewline[3em]
		\textit{lacI} & \compactfisher{16}{26}{65}{221}{S} & \compactfisher{20}{22}{61}{225}{C} & $p_{\chi^2} = 0.82$
	\end{tabular}
\end{table}

\section{Diagonal of SCA matrix instead of conservation}
In the paper we point out that the top eigenvector of the SCA matrix correlates primarily with the diagonal elements $\tilde C_{ii}$ of this matrix (or rather, with their square root), while the correlation with conservation is weaker. This is because, although both conservation and the diagonal elements of the SCA matrix can be calculated from the single-site frequencies $f_i(a)$, the relation between them is non-trivial and non-monotonic. We therefore wondered how the functional significance of the sector residues compared to that of residues that have high values of $\tilde C_{ii}$. The results can be found in Table~\ref{tab:supp_diag_results}.

\begin{table}
	\centering
	\caption{The results of the paper when replacing conservation by the diagonal of the SCA matrix.\label{tab:supp_diag_results}}
	\bigskip
	\begin{tabular}{>{\centering}p{0.15\textwidth}>{\centering}p{0.25\textwidth}>{\centering}p{0.25\textwidth}}
		Alignment & Contingency table using $\tilde C_{ii}$ & Comparison to sector ($\chi^2$ test $p$ value)
		\tabularnewline[1em]
		\toprule\tabularnewline
		PDZ & \compactfisher{12}{8}{9}{52}{C} & $p_{\chi^2} = 0.87$\tabularnewline[3em]
		DHFR & \compactfisher{13}{1}{34}{16}{C} & $p_{\chi^2} = 0.61$\tabularnewline[3em]
		potassium channels & \compactfisher{16}{21}{16}{74}{C} & $p_{\chi^2} = 0.95$\tabularnewline[3em]
		\textit{lacI} & \compactfisher{43}{27}{38}{220}{C} & $p_{\chi^2} = 0.55$
	\end{tabular}
\end{table}
	
	\section{Multiple sectors}
	\label{sec:supp_multisec}
	There are two key questions related to multiple SCA sectors: one is how to determine how many eigenvectors to analyze, and the second one is which linear combinations of eigenvectors to use for finding sectors. We briefly discussed the second question in the main text, and pointed out that while there are several approaches that have been used in the literature to find the appropriate linear combinations, these approaches have little or no theoretical motivation and have been tested only in a very limited fashion.
	
\begin{figure}
		\centering
		\includegraphics[width=0.5\textwidth]{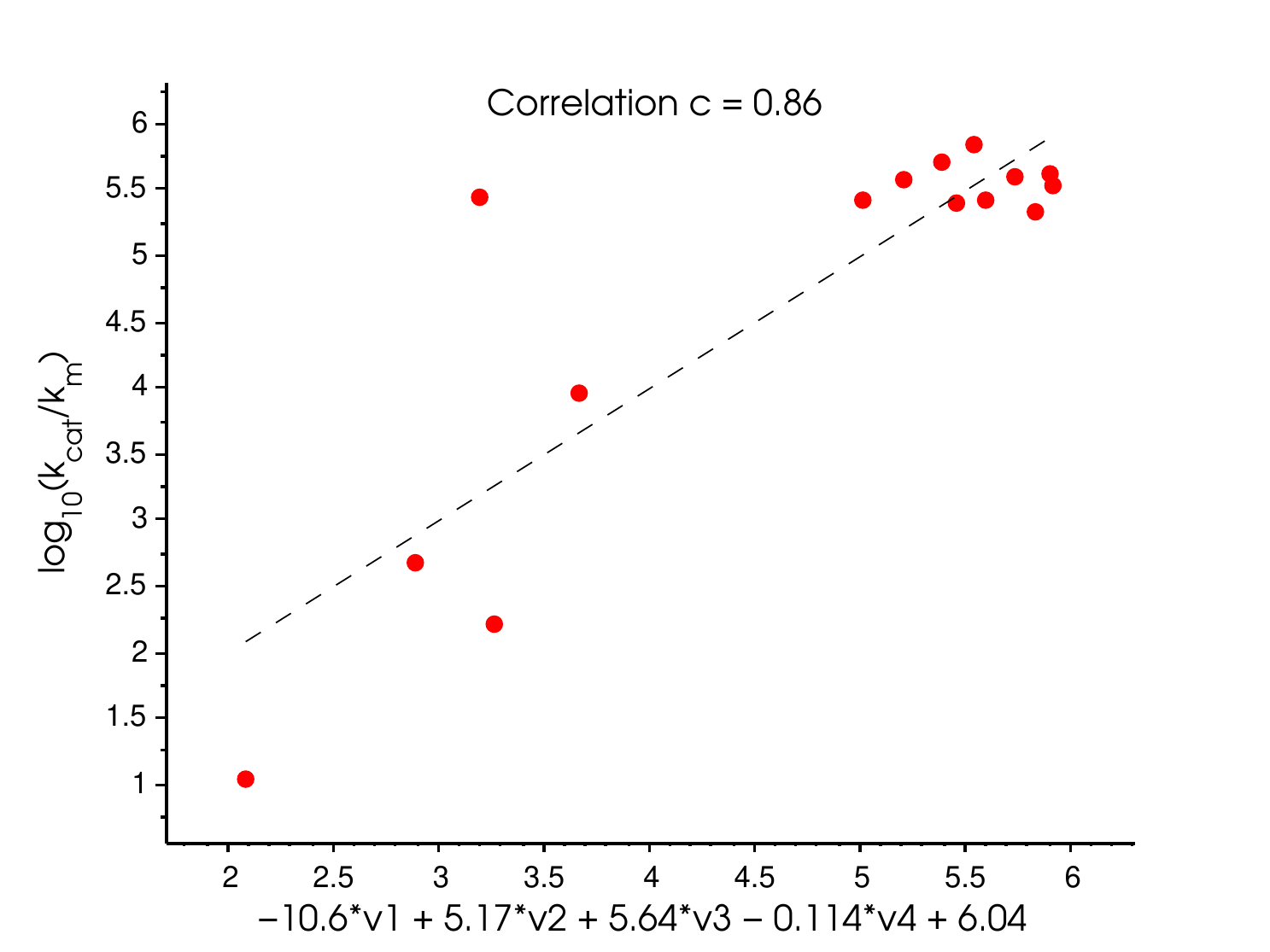}%
		\includegraphics[width=0.5\textwidth]{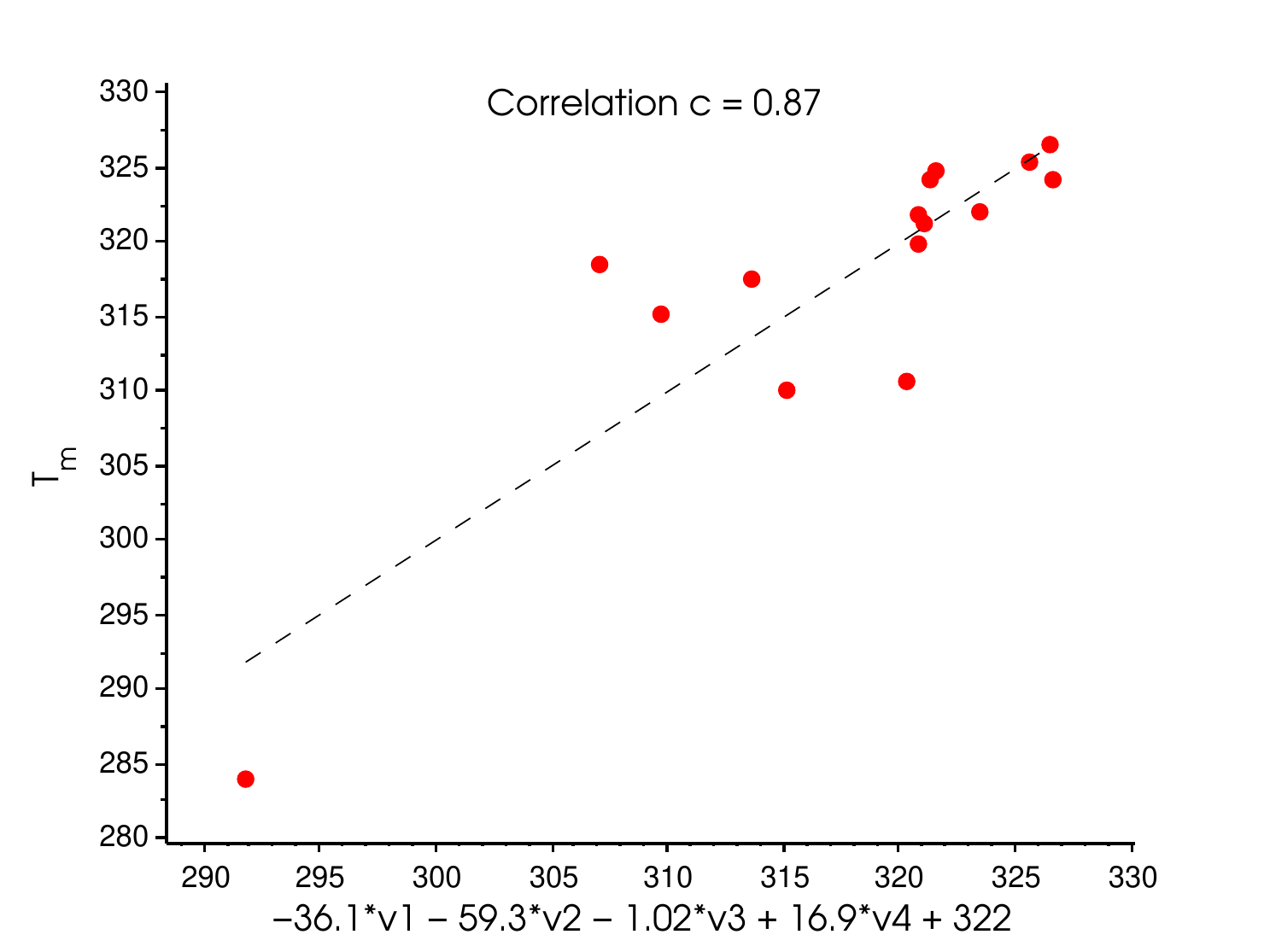}
		
		\caption{We attempt to fit binding affinity (left) or denaturation temperature (right) for the single mutants of rat trypsin described in Halabi et al.~\cite{Halabi2009} against the components of the top four eigenvectors of the SCA matrix corresponding to the mutated residues. The best linear regressions are shown on the $x$-axis. The dashed line has slope 1 and intercept 0.\label{fig:supp_lincomb_trypsin}}
	\end{figure}

\begin{figure}
		\centering
		\includegraphics[width=0.5\textwidth]{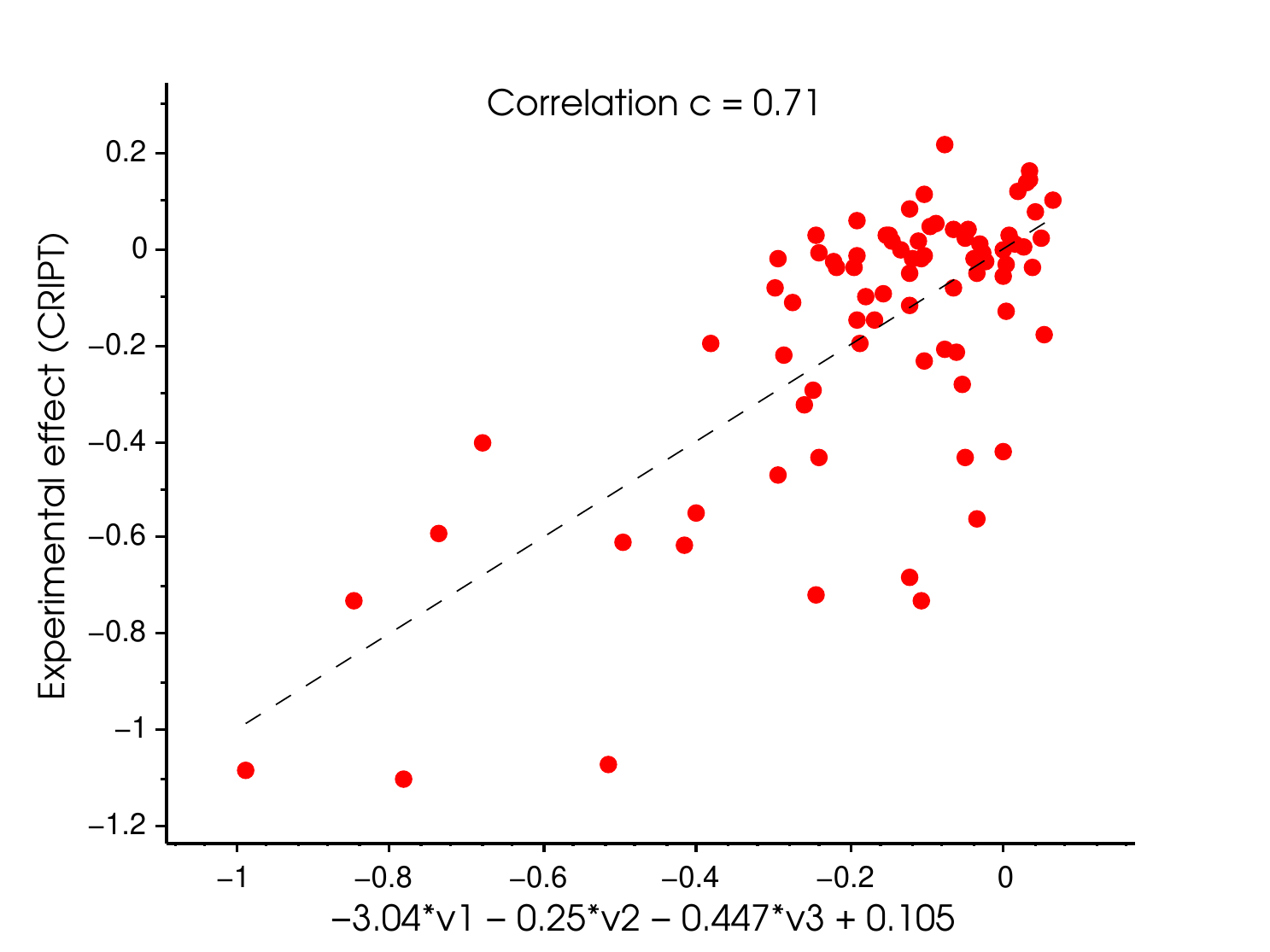}%
		\includegraphics[width=0.5\textwidth]{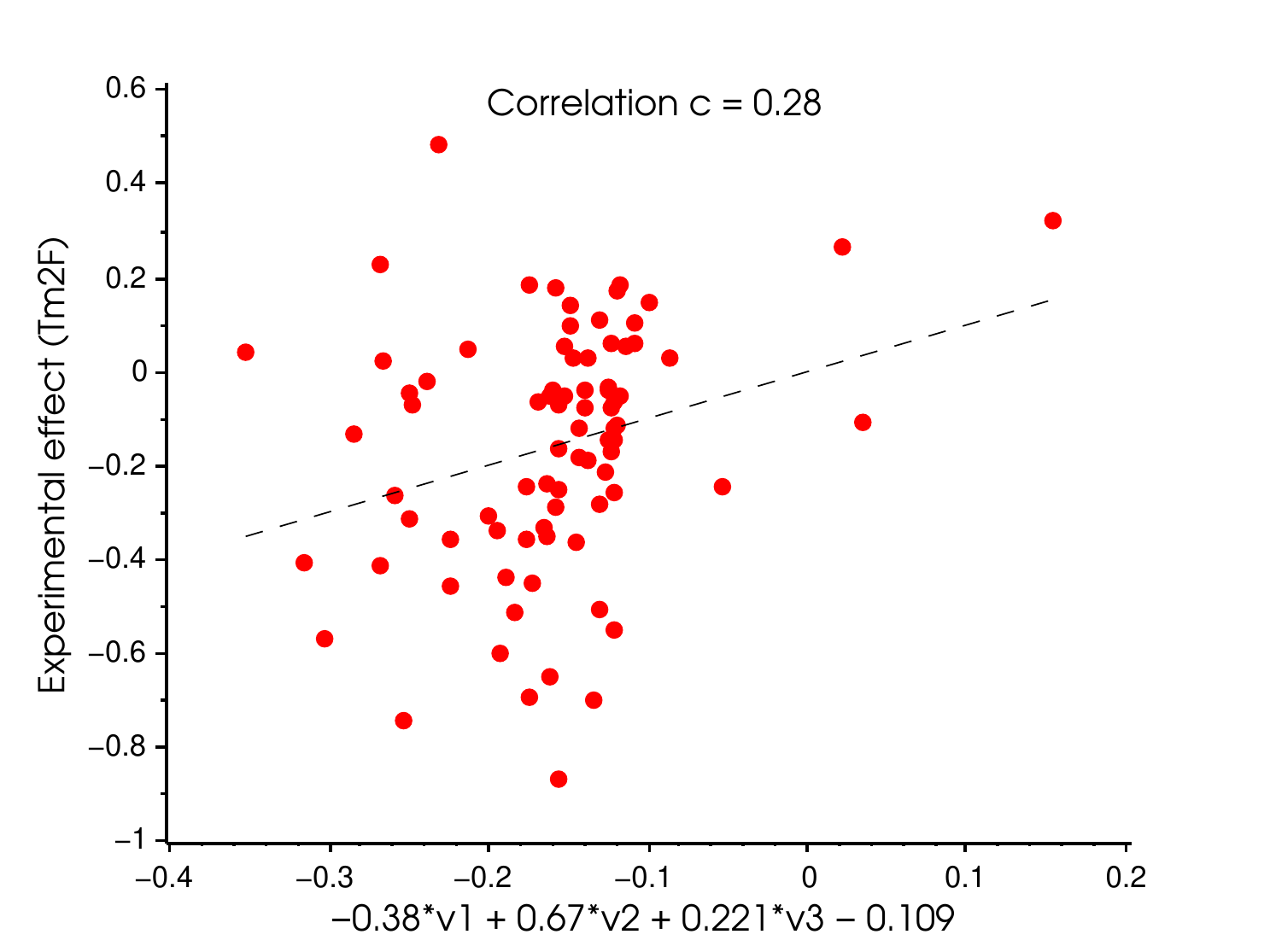}
		
		\caption{We attempt to fit the measured mutational effect for binding to the CRIPT ligand (left) or the $\text{T}_{-2}\text{F}$ ligand (right) as measured for the single mutants of \psd~described in McLaughlin Jr.\ et al.~\cite{McLaughlinJr2012} against the components of the top three eigenvectors of the SCA matrix corresponding to the mutated residues. The best linear regressions are shown on the $x$-axis. The dashed line has slope 1 and intercept 0.\label{fig:supp_lincomb_pdz}}
	\end{figure}

\begin{figure}
		\centering
		\includegraphics[width=0.5\textwidth]{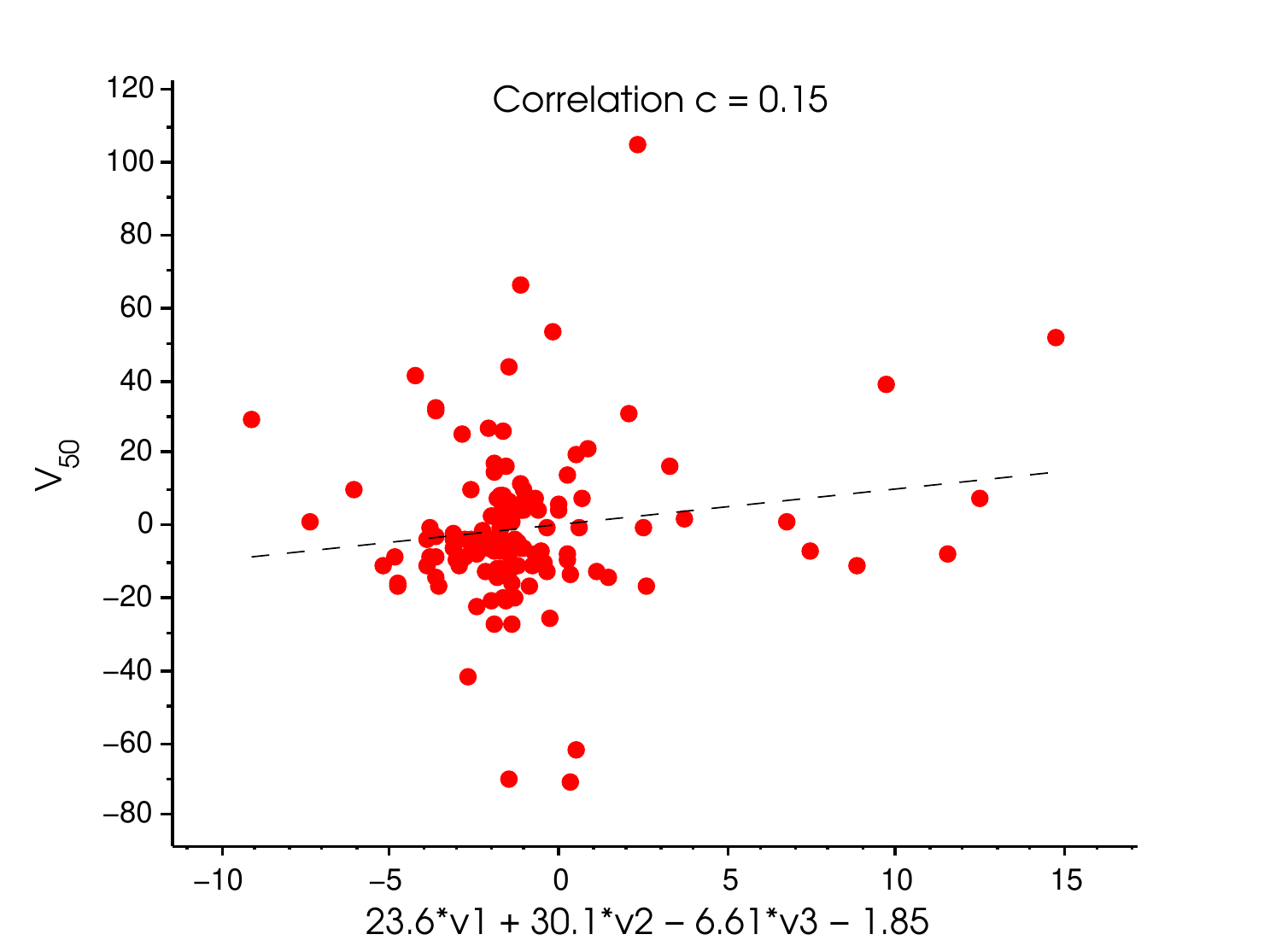}%
		\includegraphics[width=0.5\textwidth]{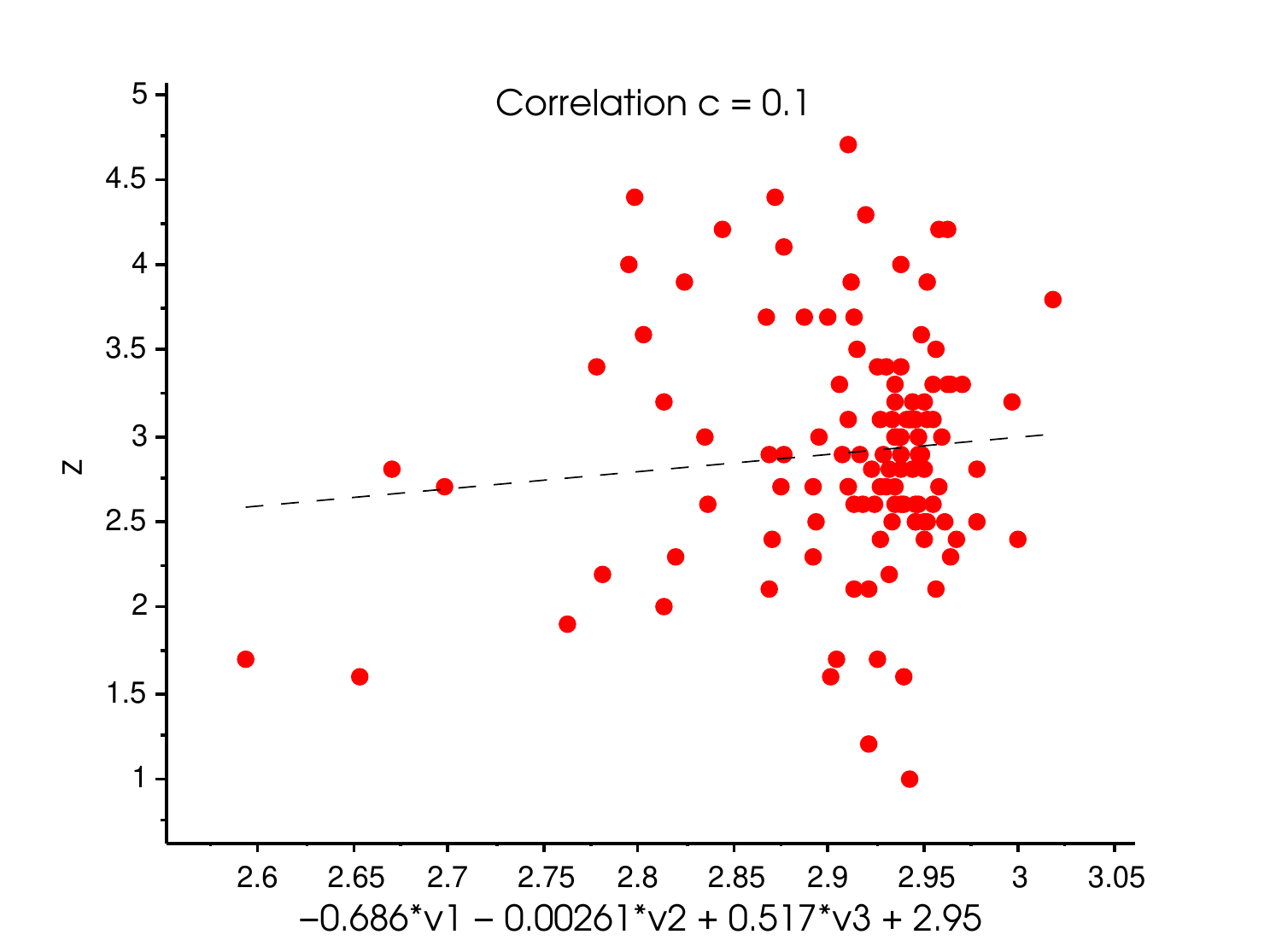}
		
		\caption{We attempt to fit the activation voltage $V_{50}$ (left) and the equivalent charge $z$  (right) measured for single mutants of the \textit{drk1} voltage-gated $K^+$ channel in Li-Smerin et al.~\cite{Li-Smerin2000} against the components of the top three eigenvectors of the SCA matrix corresponding to the mutated residues. The best linear regressions are shown on the $x$-axis. The dashed line has slope 1 and intercept 0.\label{fig:supp_lincomb_kv}}
	\end{figure}

	In the main text, we pointed out that we can instead use linear regression to find the linear combinations that best approximate the measured quantities.\footnote{This of course assumes that the relation is linear, which is far from obvious, but can be thought of as a first-order approximation.} Here we show the results of such an analysis for the case of serine protease (Figure~\ref{fig:supp_lincomb_trypsin}), PDZ (Figure~\ref{fig:supp_lincomb_pdz}), and the potassium channels (Figure~\ref{fig:supp_lincomb_kv}). While for serine protease the regression works well for both measured quantities, for PDZ we can only fit the mutational effect on binding to the CRIPT ligand, and for potassium channels the fit is not very good to either the activation potential $V_{50}$ or the equivalent charge $z$.
	
	\newcommand{\MP}{Mar\v cenko-Pastur}
	The number of eigenvectors to consider for sector determination is itself a difficult problem. It was addressed by Halabi et al.\ using an approach inspired from the analysis of financial markets~\cite{Halabi2009}. Essentially, the eigenvalue spectrum of the SCA matrices obtained for randomized alignments was obtained, and a threshold was established at the upper edge of this distribution. Any eigenvalues of the SCA matrix for the real alignment that fell below this edge were considered noise and were ignored. This approach was motivated by the \MP\ distribution, that describes the eigenvalue spectrum for a covariance matrix $C = X^T X$ associated to a random i.i.d.\ data matrix $X$. However, the assumption of identically distributed elements does not hold for alignment data, because different columns in the alignment have different amino acid distributions, even when there are no correlations between columns. For this reason, the spectrum of the SCA matrix does not resemble the \MP\ distribution, and intuitions based on this distribution need not hold in the case of SCA. In particular, eigenvalues that are below the edge obtained from randomized samples could contain perfectly valid information about the protein.
	
	Another difficulty with SCA is that the absolute value of all the elements in the covariance matrix is taken, which makes the top eigenvector become an outlier. As described in the main text, although this top eigenmode is much over the ``noise'' edge described above, the information it contains is essentially independent of correlations, and is thus indistinguishable from noise.\footnote{This is because the randomized alignments used with SCA, which are used to calculate the noise distribution, are obtained by shuffling elements in the alignment columns. This keeps the single-site frequencies intact while destroying correlations, implying that the single-site frequencies are a feature of the noise model.}

	\section{Top eigenmode of SCA matrix---some details}
	\label{sec:sca_topvec_full}
	
	Here we fill in some of the details for the model presented in the paper that can explain the correlation between the components of the top eigenvector of the SCA matrix and its diagonal entries. As in the paper, suppose we have a covariance matrix with off-diagonal entries that are biased towards positive values. The simple model we wrote for this is
	\es{eq:simpleScaMat}{M = \begin{pmatrix}
			d_1^2(1 + x) & d_1 d_2 x & \cdots & d_1 d_n x\\
			d_2 d_1 x & d_2^2(1 + x) & \cdots & d_2 d_n x\\
			\cdots & \cdots & \ddots & \cdots\\
			d_n d_1 x & d_n d_2 x & \cdots & d_n^2(1 + x)
		\end{pmatrix} \equiv \begin{pmatrix}
			\Delta_1 & d_1 d_2 x & \cdots & d_1 d_n x\\
			d_2 d_1 x & \Delta_2 & \cdots & d_2 d_n x\\
			\cdots & \cdots & \ddots & \cdots\\
			d_n d_1 x & d_n d_2 x & \cdots & \Delta_n
		\end{pmatrix}\,.}
	For simplicity, let us assume that there are no degeneracies between the $d_i$, \textit{i.e.,} that $d_i \ne d_j$ for $i \ne j$, and that $M$ is not singular, \textit{i.e.,} $d_i \ne 0$ for all $i$. Let $v = (v_1, \dotsc, v_n)$ be an eigenvector of this matrix with eigenvalue $\lambda$. Then we have
	\es{eq:evecEq}{d_i^2 v_i + d_i x \sum_j d_j v_j \stackrel != \lambda v_i\,,}
	which yields\footnote{We may worry about division by zero. Note that, from the eigenvalue equation~\eqref{eq:evecEq}, $\lambda = d_i^2$ for some $i$ if and only if $\sum_j d_j v_j = 0$ (since we assumed all $d_i \ne 0$). However, feeding this back into eq.~\eqref{eq:evecEq}, we see that this is only possible if all the $v_j$ for which $d_j^2 \ne \lambda$ are zero. Since we assumed that none of the $d_j$ vanish, $\sum_j d_j v_j = 0$ can only hold if at least two components of $v$ are non-vanishing. This, however, would imply that there is a degeneracy, which we explicitly disallowed. We thus conclude that $\lambda \ne d_i^2$ for any $i$.}
	\es{eq:evecVsDiag}{v_i = \Bigl(\sum_j d_j v_j\Bigr) \, \frac {d_i x} {\lambda - d_i^2}\,.}
	This implies that the components of the eigenvectors are related to the diagonal elements $\Delta_i = d_i^2 (1+x)$ by
	\es{eq:evecVsFctDiag}{\frac {\sqrt{\Delta_i}} {v_i} \propto \lambda - \frac {\Delta_i} {1+x}\,.}
	If we multiply eq.~\eqref{eq:evecVsDiag} by $d_i$ and sum over $i$, we can divide through by $\sum_j d_j v_j$, and get
	\es{eq:evecNormalization}{1 = x\sum_i \frac {d_i^2} {\lambda - d_i^2}\,,}
	which can be used to estimate $\lambda$. In particular, this equation allows us to show that between each consecutive pair of values $d_{i_1}$ and $d_{i_2}$, there is exactly one eigenvalue.
	
	By the Perron-Frobenius theorem, the top eigenvector can be chosen to have all components positive, and thus it should have $\lambda$ larger than all $d_i^2$. Assuming $\lambda \gg d_i^2$, which empirically seems to be the case for SCA matrices, we get an estimate for the top eigenvalue
	\es{eq:topEvalEstimate}{\lambda_{\text{top}} \approx x \sum_i d_i^2 = \frac x {1+x} \sum_i \Delta_i \equiv \frac x {1+x} \Tr M\,.}
	It should be checked that this is consistent with the condition that $\lambda$ is much larger than all $d_i^2$; this seems to be true for empirical SCA matrices. The Perron-Frobenius theorem also guarantees that all other eigenvectors of $M$ will have both positive and negative components, and therefore, according to eq.~\eqref{eq:evecVsDiag}, the corresponding eigenvalues will have to be smaller than the largest $d_i^2$. This implies that for the SCA matrices, the top eigenvector will be an outlier, \textit{i.e.,} the SCA matrices are approximately rank-1, which can indeed be observed for real alignments.
	
	Using $\lambda_{\text{top}} \gg d_i^2$ in eq.~\eqref{eq:evecVsDiag}, we get
	\es{eq:evecVsSqrtDiag}{v_{i,\text{top}} \approx \Bigl(\sum_j d_j v_{j, \text{top}}\Bigr) \frac x {\lambda_{\text{top}}} \times \sqrt{\Delta_i}\,,}
	which is the observed linear relation between the top eigenvector and the square root of the diagonal elements of the SCA matrix. This argument shows that the top eigenvector is strongly correlated with single-site statistics and thus largely independent of correlations between positions. It is important to emphasize that this does not mean that there is no information contained in this mode, but only that most of this information can be obtained without any analysis of correlations.
	
	As mentioned in the paper, we emphasize again that in this derivation the origin of the off-diagonal entries is not specified. They could be an artifact of sampling noise, they could come from actual non-specific correlations between positions, or they could be due to a non-trivial phylogenetic structure of the alignment, as suggested by Halabi et al.~\cite{Halabi2009}.
	
\section{Data availability}
This study is based on several alignments that are either directly accessible online or upon request from the authors of earlier studies. Some of the alignments can also be created using publicly-available software. This is described in detail below:
\begin{enumerate}
 	\item \hhblits alignments for PDZ, DHFR, \textit{lacI}, and the $K^+$ channels were generated as described in sections~\ref{sec:methods_alignments} and~\ref{sec:supp_alignments}. \hhblits can be obtained from \url{ftp://toolkit.genzentrum.lmu.de/pub/HH-suite/}.

	\item \pfam alignments were downloaded from \url{http://pfam.xfam.org/}.
	
	\item The PDZ and DHFR alignments used in McLaughlin Jr.\ et al.\ 2012~\cite{McLaughlinJr2012} and Reynolds et al.\ 2011\cite{Reynolds2011}, respectively, were obtained upon request from the Ranganathan lab, \url{http://systems.swmed.edu/rr_lab/index.html}.

	\item The potassium channels alignment from Lee et al.\ 2009~\cite{Lee2009} is available as supporting information to that article, \url{http://www.plosbiology.org/article/info:doi/10.1371/journal.pbio.1000047#s5}.
\end{enumerate}

The mutagenesis data that we used is available from the following sources:
\begin{enumerate}
	\item The PDZ mutation data is available directly from the Ranganathan lab website, \url{http://systems.swmed.edu/rr_lab/papers/McLaughlin_Ranganathan/McLaughlin_Ranganathan_data.mat.zip}.
	
	\item The DHFR light-sensitivity data was obtained upon request from the Ranganathan lab.
	
	\item The $K^+$ channels data we used is available in the main text of Li-Smerin et al.\ 2000~\cite{Li-Smerin2000}.
	
	\item The \textit{lacI} data from Markiewicz et al.\ 1994~\cite{Markiewicz1994} is available for download at \url{http://sift.bii.a-star.edu.sg/www/test_sets/LacITable.txt}.
\end{enumerate}

The raw alignment and mutagenesis data were further processed using custom-made {\sffamily Matlab} and {\sffamily C++} code, publicly available for download from \url{https://bitbucket.org/ttesileanu/multicov}.

\section{Differences from previous version}
There are several differences between this version and the previous one uploaded to the \arxiv. The most important difference is that we decided to use a consistent SCA method and a consistent methodology for generating the alignments for all the datasets we consider.\footnote{This was suggested to us by a reviewer.} Previously we were using the SCA method and alignments from the earlier studies that had compared SCA to mutational data. We also fixed a bug that caused a small error in the conservation values.

\newpage
\bibliography{bibliography,extrabib}

\begin{thebibliography}{10}
\providecommand{\url}[1]{\texttt{#1}}
\providecommand{\urlprefix}{URL }
\expandafter\ifx\csname urlstyle\endcsname\relax
  \providecommand{\doi}[1]{doi:\discretionary{}{}{}#1}\else
  \providecommand{\doi}{doi:\discretionary{}{}{}\begingroup
  \urlstyle{rm}\Url}\fi
\providecommand{\bibAnnoteFile}[1]{%
  \IfFileExists{#1}{\begin{quotation}\noindent\textsc{Key:} #1\\
  \textsc{Annotation:}\ \input{#1}\end{quotation}}{}}
\providecommand{\bibAnnote}[2]{%
  \begin{quotation}\noindent\textsc{Key:} #1\\
  \textsc{Annotation:}\ #2\end{quotation}}
\providecommand{\eprint}[2][]{\url{#2}}

\bibitem{Do2008}
Do CB, Katoh K (2008) {Protein multiple sequence alignment}.
\newblock Methods in Molecular Biology 484: 379--413.
\bibAnnoteFile{Do2008}

\bibitem{Notredame2002}
Notredame C (2002) {Recent progress in multiple sequence alignment: a survey}.
\newblock Pharmacogenomics 3: 131--144.
\bibAnnoteFile{Notredame2002}

\bibitem{Schneider1986}
Schneider TD, Stormo GD, Gold L, Ehrenfeucht A (1986) {Information content of
  binding sites on nucleotide sequences}.
\newblock Journal of Molecular Biology 188: 415--431.
\bibAnnoteFile{Schneider1986}

\bibitem{Zvelebil1987}
Zvelebil MJ, Barton GJ, Taylor WR, Sternberg MJE (1987) {Prediction of protein
  secondary structure and active sites using the alignment of homologous
  sequences}.
\newblock Journal of Molecular Biology 195: 957--961.
\bibAnnoteFile{Zvelebil1987}

\bibitem{Hollstein1991}
Hollstein M, Sidransky D, Vogelstein B, Harris CC (1991) {p53 Mutations in
  Human Cancers}.
\newblock Science 253: 49--53.
\bibAnnoteFile{Hollstein1991}

\bibitem{Cargill1999}
Cargill M, Altshuler D, Ireland J, Sklar P, Ardlie K, et~al. (1999)
  {Characterization of single-nucleotide polymorphisms in coding regions of
  human genes}.
\newblock Nature Genetics 22: 231--238.
\bibAnnoteFile{Cargill1999}

\bibitem{Ng2003}
Ng PC, Henikoff S (2003) {SIFT: predicting amino acid changes that affect
  protein function}.
\newblock Nucleic Acids Research 31: 3812--3814.
\bibAnnoteFile{Ng2003}

\bibitem{Russ2005}
Russ WP, Lowery DM, Mishra P, Yaffe MB, Ranganathan R (2005) {Natural-like
  function in artificial WW domains}.
\newblock Nature 437: 579--583.
\bibAnnoteFile{Russ2005}

\bibitem{Socolich2005}
Socolich M, Lockless SW, Russ WP, Lee H, Gardner KH, et~al. (2005)
  {Evolutionary information for specifying a protein fold}.
\newblock Nature 437: 512--518.
\bibAnnoteFile{Socolich2005}

\bibitem{Marks}
Marks DS, Colwell LJ, Sheridan R, Hopf TA, Pagnani A, et~al. (2011) {Protein 3D
  Structure Computed from Evolutionary Sequence Variation}.
\newblock PLoS ONE 6: e28766.
\bibAnnoteFile{Marks}

\bibitem{Morcos2011}
Morcos F, Pagnani A, Lunt B, Bertolino A, Marks DS, et~al. (2011)
  {Direct-coupling analysis of residue co-evolution captures native contacts
  across many protein families}.
\newblock PNAS 108: E1293--E1301.
\bibAnnoteFile{Morcos2011}

\bibitem{Weigt2009}
Weigt M, White RA, Szurmant H, Hoch JA, Hwa T (2009) {Identification of direct
  residue contacts in protein-protein interaction by message passing}.
\newblock PNAS 106: 67--72.
\bibAnnoteFile{Weigt2009}

\bibitem{Lockless1999}
Lockless SW, Ranganathan R (1999) {Evolutionarily Conserved Pathways of
  Energetic Connectivity in Protein Families}.
\newblock Science 286: 295--299.
\bibAnnoteFile{Lockless1999}

\bibitem{Fuentes2004}
Fuentes EJ, Der CJ, Lee AL (2004) {Ligand-dependent dynamics and intramolecular
  signaling in a PDZ domain}.
\newblock Journal of Molecular Biology 335: 1105--1115.
\bibAnnoteFile{Fuentes2004}

\bibitem{Peterson2004}
Peterson FC, Penkert RR, Volkman BF, Prehoda KE (2004) {Cdc42 regulates the
  Par-6 PDZ domain through an allosteric CRIB-PDZ transition}.
\newblock Molecular Cell 13: 665--676.
\bibAnnoteFile{Peterson2004}

\bibitem{Suel2003}
S\"{u}el GM, Lockless SW, Wall MA, Ranganathan R (2003) {Evolutionarily
  conserved networks of residues mediate allosteric communication in proteins}.
\newblock Nature 10: 59--69.
\bibAnnoteFile{Suel2003}

\bibitem{Hatley2003}
Hatley ME, Lockless SW, Gibson SK, Gilman AG, Ranganathan R (2003) {Allosteric
  determinants in guanine nucleotide-binding proteins.}
\newblock PNAS 100: 14445--14450.
\bibAnnoteFile{Hatley2003}

\bibitem{Halabi2009}
Halabi N, Rivoire O, Leibler S, Ranganathan R (2009) {Protein sectors:
  evolutionary units of three-dimensional structure}.
\newblock Cell 138: 774--786.
\bibAnnoteFile{Halabi2009}

\bibitem{Ranganathan2011url}
Ranganathan R, Rivoire O (2012).
\newblock {Note 109: A summary of SCA calculations}.
\newblock Available online at
  \url{http://systems.swmed.edu/rr\_lab/Note109\_files/Note109\_v3.pdf}.
\bibAnnoteFile{Ranganathan2011url}

\bibitem{Smock2010}
Smock RG, Rivoire O, Russ WP, Swain JF, Leibler S, et~al. (2010) {An
  interdomain sector mediating allostery in Hsp70 molecular chaperones.}
\newblock Molecular Systems Biology 6: 414.
\bibAnnoteFile{Smock2010}

\bibitem{Bouchaud2009}
Bouchaud JP, Potters M (2009) {Financial Applications of Random Matrix Theory:
  a short review}.
\newblock arXiv 0910.1205.
\bibAnnoteFile{Bouchaud2009}

\bibitem{Rivoire2012}
Rivoire O (2013) {Elements of Coevolution in Biological Sequences}.
\newblock Physical Review Letters 110: 178102.
\bibAnnoteFile{Rivoire2012}

\bibitem{Colwell2014}
Colwell LJ, Brenner MP, Murray AW (2014) {Conservation weighting functions
  enable covariance analyses to detect functionally important amino acids.}
\newblock PLoS ONE 9: e107723.
\bibAnnoteFile{Colwell2014}

\bibitem{Reynolds2011}
Reynolds KA, {McLaughlin Jr} RN, Ranganathan R (2011) {Hot spots for allosteric
  regulation on protein surfaces}.
\newblock Cell 147: 1564--75.
\bibAnnoteFile{Reynolds2011}

\bibitem{McLaughlinJr2012}
{McLaughlin Jr} RN, Poelwijk FJ, Raman A, Gosal WS, Ranganathan R (2012) {The
  spatial architecture of protein function and adaptation}.
\newblock Nature 491: 138--142.
\bibAnnoteFile{McLaughlinJr2012}

\bibitem{Kern2003}
Kern D, Zuiderweg ERP (2003) {The role of dynamics in allosteric regulation}.
\newblock Current Opinion in Structural Biology 13: 748--757.
\bibAnnoteFile{Kern2003}

\bibitem{Matoba2003}
Matoba Y, Sugiyama M (2003) {Atomic resolution structure of prokaryotic
  phospholipase A2: analysis of internal motion and implication for a catalytic
  mechanism}.
\newblock Proteins 51: 453--69.
\bibAnnoteFile{Matoba2003}

\bibitem{Fraser2009}
Fraser JS, Clarkson MW, Degnan SC, Erion R, Kern D, et~al. (2009) {Hidden
  alternative structures of proline isomerase essential for catalysis}.
\newblock Nature 462: 669--673.
\bibAnnoteFile{Fraser2009}

\bibitem{Dhulesia2008}
Dhulesia A, Gsponer J, Vendruscolo M (2008) {Mapping of two networks of
  residues that exhibit structural and dynamical changes upon binding in a PDZ
  domain protein}.
\newblock Journal of the American Chemical Society 130: 8931--8939.
\bibAnnoteFile{Dhulesia2008}

\bibitem{Remmert2012}
Remmert M, Biegert A, Hauser A, S\"{o}ding J (2012) {HHblits: lightning-fast
  iterative protein sequence searching by HMM-HMM alignment}.
\newblock Nature Methods 9: 173--175.
\bibAnnoteFile{Remmert2012}

\bibitem{Lee2009}
Lee SY, Banerjee A, MacKinnon R (2009) {Two separate interfaces between the
  voltage sensor and pore are required for the function of voltage-dependent
  K(+) channels}.
\newblock PLoS Biology 7: 676--686.
\bibAnnoteFile{Lee2009}

\bibitem{Mann1947}
Mann HB, Whitney DR (1947) {On a test of whether one of two random variables is
  stochastically larger than the other}.
\newblock Annals of Mathematical Statistics 18: 50--60.
\bibAnnoteFile{Mann1947}

\bibitem{Li-Smerin2000}
Li-Smerin Y, Hackos DH, Swartz KJ (2000) {Alpha-helical structural elements
  within the voltage-sensing domains of a K(+) channel}.
\newblock The Journal of General Physiology 115: 33--49.
\bibAnnoteFile{Li-Smerin2000}

\bibitem{Markiewicz1994}
Markiewicz P, Kleina LG, Cruz C, Ehret S, Miller JH (1994) {Genetic Studies of
  the lac Repressor -- XIV. Analysis of 4000 Altered Escherichia coli lac
  Repressors Reveals Essential and Non-essential Residues, as well as
  ``Spacers'' which do not Require a Specific Sequence}.
\newblock Journal of Molecular Biology 240: 421--433.
\bibAnnoteFile{Markiewicz1994}

\bibitem{Mirny1999}
Mirny LA, Shakhnovich EI (1999) {Universally Conserved Positions in Protein
  Folds: Reading Evolutionary Signals about Stability, Folding Kinetics and
  Function}.
\newblock Journal of Molecular Biology 291: 177--196.
\bibAnnoteFile{Mirny1999}

\bibitem{Fodor2004}
Fodor AA, Aldrich RW (2004) {Influence of conservation on calculations of amino
  acid covariance in multiple sequence alignments}.
\newblock Proteins 56: 211--21.
\bibAnnoteFile{Fodor2004}

\bibitem{Dunn2008}
Dunn SD, Wahl LM, Gloor GB (2008) {Mutual information without the influence of
  phylogeny or entropy dramatically improves residue contact prediction.}
\newblock Bioinformatics 24: 333--40.
\bibAnnoteFile{Dunn2008}

\bibitem{Freedman1981}
Freedman D, Diaconis P (1981) {On the histogram as a density estimator: L2
  theory}.
\newblock Zeitschrift fuer Wahrscheinlichkeitstheorie und Verwandte Gebiete 57:
  453--476.
\bibAnnoteFile{Freedman1981}

\end{thebibliography}

\end{document}